\documentclass[twocolumn]{article}  
\usepackage[margin=0.6in,columnsep=0.15in]{geometry}
\usepackage[utf8]{inputenc}

\usepackage[usenames,dvipsnames]{xcolor}
\PassOptionsToPackage{hyphens}{url}
\usepackage[colorlinks,citecolor=Gray,urlcolor=Gray,linkcolor=Blue]{hyperref}

\usepackage{graphicx}
\usepackage{mathtools}
\usepackage{booktabs}
\usepackage{tabularx}
\usepackage{tikz}
\usepackage[multi-part-units=single]{siunitx}
\usepackage{dcolumn}
\usepackage{multirow}
\usepackage{nidanfloat}
\usepackage{braket}
\usepackage{tcolorbox}
\usepackage[capitalize]{cleveref}
\usepackage{bm}
\usepackage[version=4]{mhchem}

\newlength{\normalparindent}
\AtBeginDocument{\setlength{\normalparindent}{\parindent}}
\tcbset{before upper={\setlength{\parindent}{\normalparindent}}}

\newtcolorbox[auto counter]{mybox}[2][]{floatplacement=t!,float,fonttitle=\bfseries,title=Box~\thetcbcounter\ | #2,#1}

\usepackage[lcgreekalpha,upint]{stix}
\usepackage[scaled=0.8]{FiraMono}

\newcommand{\eqlabelleft}{(} 
\newcommand{\eqlabelright}{)} 
\DeclareRobustCommand{\pcref}[1]{%
\begingroup%
\renewcommand{\eqlabelleft}{}%
\renewcommand{\eqlabelright}{}%
\cref{#1}%
\endgroup%
}
\creflabelformat{equation}{\eqlabelleft#2#1#3\eqlabelright}

\usepackage[tiny]{titlesec}

\usepackage[inline]{enumitem}
\setlist{nosep}

\usepackage[format=plain]{caption}
\DeclareCaptionLabelSeparator{pipe}{ $|$ }
\usepackage{authblk}

\usepackage{ifpdf}
\ifpdf%
\usepackage[style=authoryear-comp]{biblatex}
\else
\usepackage[style=authoryear-comp,backend=bibtex]{biblatex}
\fi
\let\citep\autocite%
\let\citet\textcite%
\AtBeginBibliography{\footnotesize}

\date{}

\makeatletter
\def\blfootnote{\xdef\@thefnmark{}\@footnotetext}
\makeatother

\title{\Large Ab-initio quantum chemistry with neural-network wavefunctions}

\author[1,$\dagger$]{Jan Hermann}
\author[2,$\dagger$]{James Spencer}
\author[3,4,$\dagger$]{Kenny Choo}
\author[5]{Antonio Mezzacapo}
\author[6]{W. M. C. Foulkes}
\author[2,6,*]{David Pfau}
\author[7,*]{Giuseppe Carleo}
\author[1,8,9,10,*]{Frank Noé}
\affil[1]{FU Berlin, Department of Mathematics and Computer Science, Arnimallee 12, 14195 Berlin, Germany}
\affil[2]{DeepMind, 6 Pancras Square, London N1C 4AG, United Kingdom}
\affil[3]{University of Zurich, Department of Physics, Winterthurerstrasse 190, 8057 Zurich, Switzerland}
\affil[4]{IBM Quantum, IBM Research Zurich, Saumerstrasse 4, 8803 Ruschlikon, Switzerland}
\affil[5]{IBM Quantum, Thomas J. Watson Research Center, Yorktown Heights, New York 10598, USA}
\affil[6]{Imperial College London, Department of Physics, South Kensington Campus, London SW7 2AZ, United Kingdom}
\affil[7]{EPFL, Institute of Physics, CH-1015 Lausanne, Switzerland}
\affil[8]{FU Berlin, Department of Physics, Arnimallee 12, 14195 Berlin, Germany}
\affil[9]{Rice University, Department of Chemistry, 6100 Main St, Houston, TX 77005, United States}
\affil[10]{Microsoft Research Cambridge, 21 Station Rd, Cambridge CB1 2FB, United Kingdom}

\begin{document}

\hyphenation{Schrö-din-ger}

\twocolumn[{%
  \maketitle
  \vspace{-3em}
  \begin{center}
  \begin{minipage}{0.85\linewidth}
    \small
    \paragraph{Abstract}
    Machine learning and specifically deep-learning methods have outperformed human capabilities in many pattern recognition and data processing problems, in game playing, and now also play an increasingly important role in scientific discovery.
    A key application of machine learning in the molecular sciences is to learn potential energy surfaces or force fields from ab-initio solutions of the electronic Schrödinger equation using datasets obtained with density functional theory, coupled cluster, or other quantum chemistry methods.
    Here we review a recent and complementary approach: using machine learning to aid the direct solution of quantum chemistry problems from first principles.
    Specifically, we focus on quantum Monte Carlo (QMC) methods that use neural network ansatz functions in order to solve the electronic Schrödinger equation, both in first and second quantization, computing ground and excited states, and generalizing over multiple nuclear configurations.
    Compared to existing quantum chemistry methods, these new deep QMC methods have the potential to generate highly accurate solutions of the Schrödinger equation at relatively modest computational cost.
  \end{minipage}
  \end{center}
  \vspace{1em}
}]

\blfootnote{$^\dagger$These authors contributed equally}
\blfootnote{\hangindent=3em\raggedright%
$^*$Emails: frank.noe@fu-berlin.de, giuseppe.carleo@epfl.ch, pfau@google.com}


\section{Introduction}
\label{sec:intro}

In the past decade, machine learning (ML) has made inroads into many areas of the physical sciences \citep{CarleoRMP19}, often outperforming more traditional computational methods \citep{jumper2021highly,DeringerN21} or offering entirely new approaches to solve scientific problems \citep{NoeS19,HuangNC20}.
Quantum chemistry (QC) has been among the first fields to have been affected by this revolution \citep{TkatchenkoNC20,vonLilienfeldNC20,NoeARPC20}.
Most applications of ML in QC have been concerned with supervised learning of molecular properties from molecular structure \citep{DralJPCL20}, either across conformational \citep{UnkeCR21} or chemical space \citep{vonLilienfeldNRC20}, as well as with unsupervised learning for the generation of novel molecules \citep{BianJMM21}.
These methods all require a pre-existing dataset of molecules and their properties as an input, typically obtained with standard methods of QC such as density functional theory \citep{JonesDFTReview20215} or coupled cluster theory \citep{Bartlett2007}.
In these scenarios, ML accurately approximates a given method of QC at vastly increased computational efficiency.
This approach has been already reviewed in other works cited above.
In contrast, the current review focuses on the complementary use of ML as an ab-initio technique in QC, which requires no external data and instead recovers molecular properties from first principles.
Here, ML is ``integrated'' into QC, with the goal of arriving at ab-initio methods with a more favourable accuracy--efficiency trade-off than traditional QC methods.

\begin{figure}[t]
\centering
\begin{tikzpicture}
\node[below right, inner sep=0pt] at (0.5,0) {\includegraphics[width=0.85\linewidth]{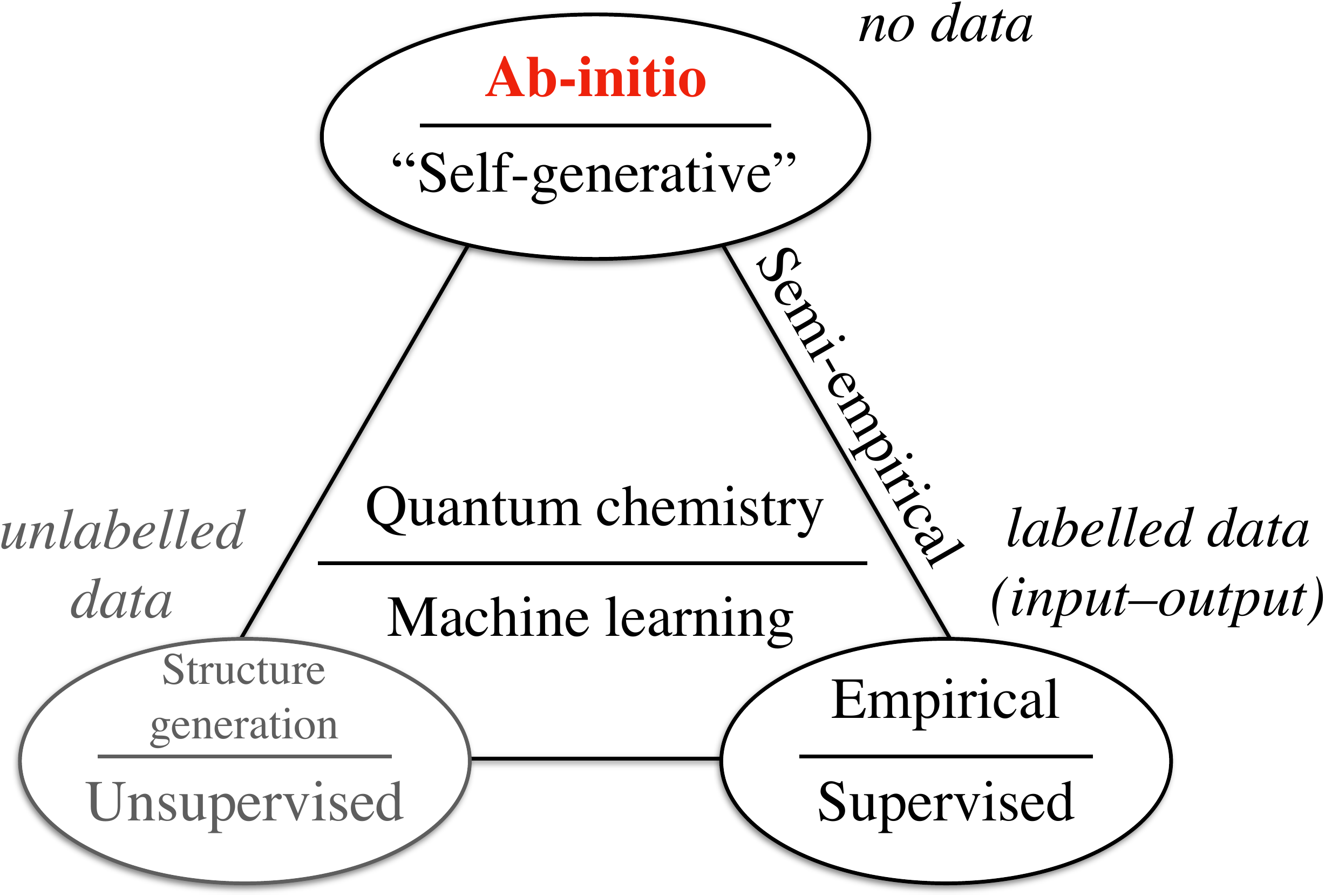}};
\node[below right] at (0,0) {\bfseries a};
\node[below right, inner sep=0pt] at (0.5,-5.5) {\includegraphics[width=0.9\columnwidth]{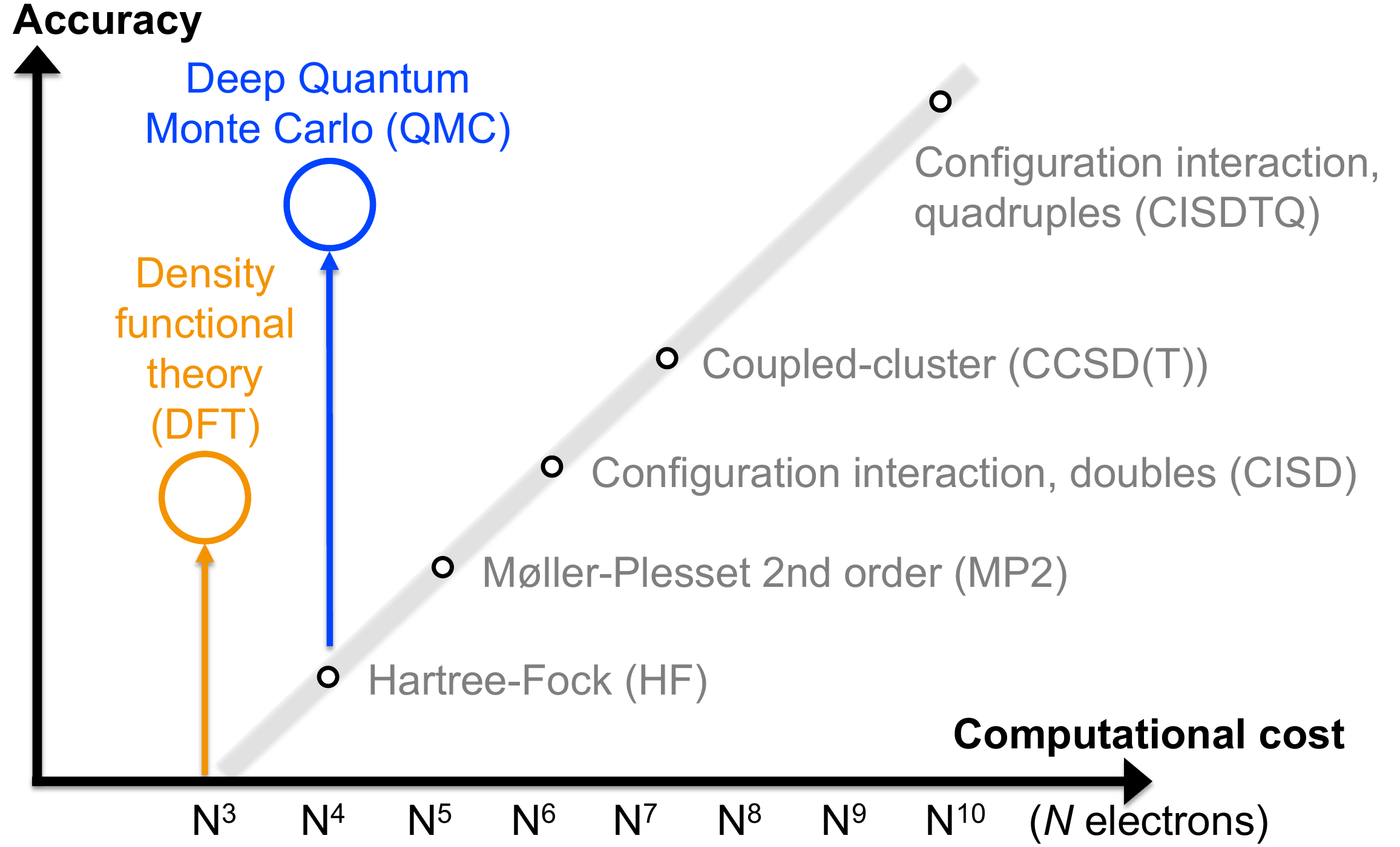}};
\node[below right] at (0,-5.5) {\bfseries b};
\end{tikzpicture}
\caption{\textbf{Quantum chemistry and machine learning.}
(\textbf a) Machine learning disciplines and their dependence on data can be mapped to disciplines in quantum chemistry.
This work reviews the use of machine learning in ab-initio quantum chemistry, where the only input to machine learning is the Schrödinger equation itself.
This approach uses self-generated data, rather than relying on external data.
The closest analogue in machine learning is reinforcement learning with self-play, which substitutes data from an external environment with data generated by the agent, though in many other respects the two approaches are distinct.
(\textbf b) Trade-off between computational efficiency and accuracy in quantum chemistry methods.
Accuracy of electronic structure methods against the asymptotic scaling of their computational cost with system size, $N$.
Popular methods, such as density functional theory, are outliers from the general trend.
}\label{fig:QC}
\end{figure}

The goal of computational chemistry is to predict properties of known molecules and to design molecules with desired properties.
Most molecular properties are determined by the behaviour of the electrons, so QC methods attempt to approximate the Schrödinger equation for electrons in molecules.
Traditionally, QC methods are divided into ab-initio and semi-empirical methods, where the former have no fitted parameters determined from external data, whereas the latter do.
Methods that do not use quantum mechanics at all (such as force fields) are called empirical and are typically not considered part of QC, although this view may be changing with the advent of principled and accurate ML-based empirical methods.
It is useful to cast these three categories of methods in the light of ML terminology (\cref{fig:QC}a).

ML can be roughly divided into supervised, unsupervised, and reinforcement learning.
In supervised learning the ML model learns to predict the labels (outputs) of the data (inputs) from a given dataset so as to minimize the difference between the predicted and reference labels.
By identifying the inputs with molecular structures and the outputs with molecular properties, all semi-empirical and empirical methods of QC fit into supervised learning, but using mostly relatively simple and physically motivated functional forms rather than the more general and highly flexible functions typical for ML\@.
Vice versa, the many recent successful supervised ML models that predict energies or other molecular properties based on QC training data can be classified as empirical methods \citep{DeringerCR21,BehlerCR21,UnkeCR21,MusilCR21}.
Unsupervised learning is concerned with unlabelled data, and the general task is to learn the underlying probability distribution that would generate a given dataset.
Examples in chemistry include generative models for structural formulas \citep{Gomez-BombarelliACS18} as well as full 3D structures of molecules \citep{NoeS19,Hoogeboom22}, and in physics the estimation of quantum states from measurements, known as quantum tomography \citep{TorlaiNP18}.
Finally, in reinforcement learning, the ML model (also referred to as an \emph{agent} is able to interact directly with its environment, rather than to just passively receive data.
Here, the aim is for the agent to learn a \emph{policy} for how to interact with the environment so as to maximize a long-term reward \citep{sutton2018reinforcement}.
Reinforcement learning is behind some of the most prominent successes of ML such as playing games at a superhuman level \citep{tesauro1994td, mnih2015human, silver2016mastering} or the control of plasma in tokamaks \citep{DegraveN22}.
In certain settings the agent can self-generate data by treating its own policy as the environment.
This is known as {\em self-play}, and has been the basis for many advances in symmetric games \citep{heinrich2015fictitious, SilverS18}.
Although there are many key differences, this is the branch of ML conceptually most similar to ab-initio QC, in the sense that no external data other than the rules of the system or game are required for either.

In the traditional picture, one moves from empirical to ab-initio methods by retaining more of the first-principles physics.
Similarly, there is a general trend for ML models in chemistry to encode an increasing amount of molecular physics.
This includes physical constraints such as energy conservation \citep{ChmielaSA17}, invariance and equivariance of molecular properties with respect to rotation, translation, or exchange of indistinguishable particles \citep{BehlerPRL07,SchuttNC17}, as well as other physical concepts such as many-body expansions \citep{DrautzPRB19} or even surrogate quantum-mechanical models \citep{LiJCTC18a,SchuttNC19,KirkpatrickS21}.
Similar considerations can be made for the problem of ab-initio learning of solutions to the electronic Schrödinger equation introduced here and we will discuss different strategies throughout the review.

The Schrödinger equation is an eigenvalue problem that can be equivalently formulated via several variational principles---its solutions, the \emph{eigenstate} wavefunctions and energies, can be found by searching for stationary points of certain functionals over the space of all physically admissible wavefunctions.
Importantly, the ground state of a molecule can be found by minimizing the energy expectation value of a wavefunction.
This principle underlies many ab-initio QC methods, and also the methods in this review, as such a variational principle naturally defines a ML problem---the eigenstates (such as the ground state) are represented as a neural network and the parameters of that network are obtained by minimizing the variational electronic energy.
The reviewed methods differ in the particular form of the neural-network ansatz used, as described below.

\begin{figure*}[t]
\centering
\includegraphics[width=\columnwidth-0.5em]{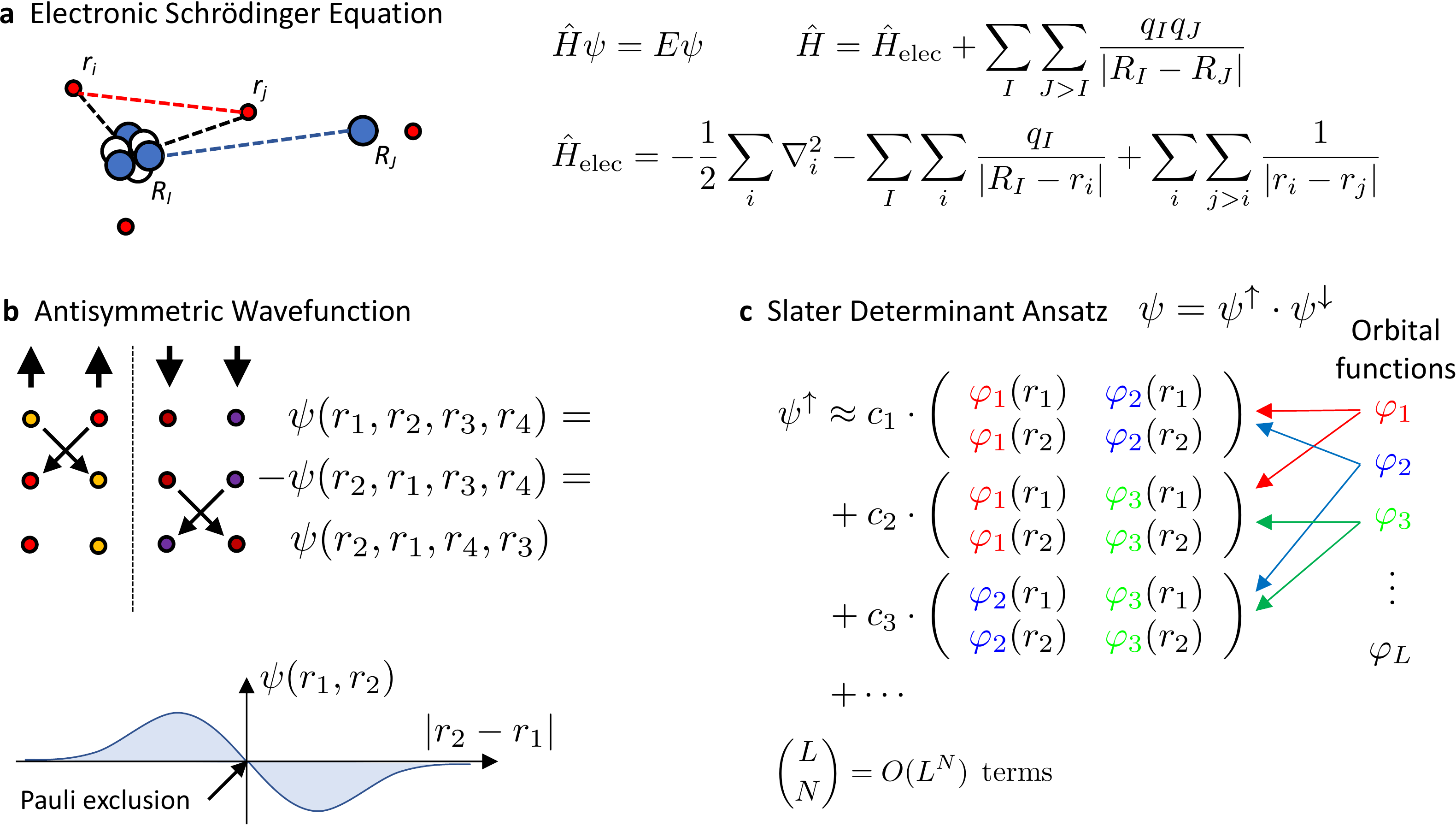}\hspace{1em}
\includegraphics[width=\columnwidth-0.5em]{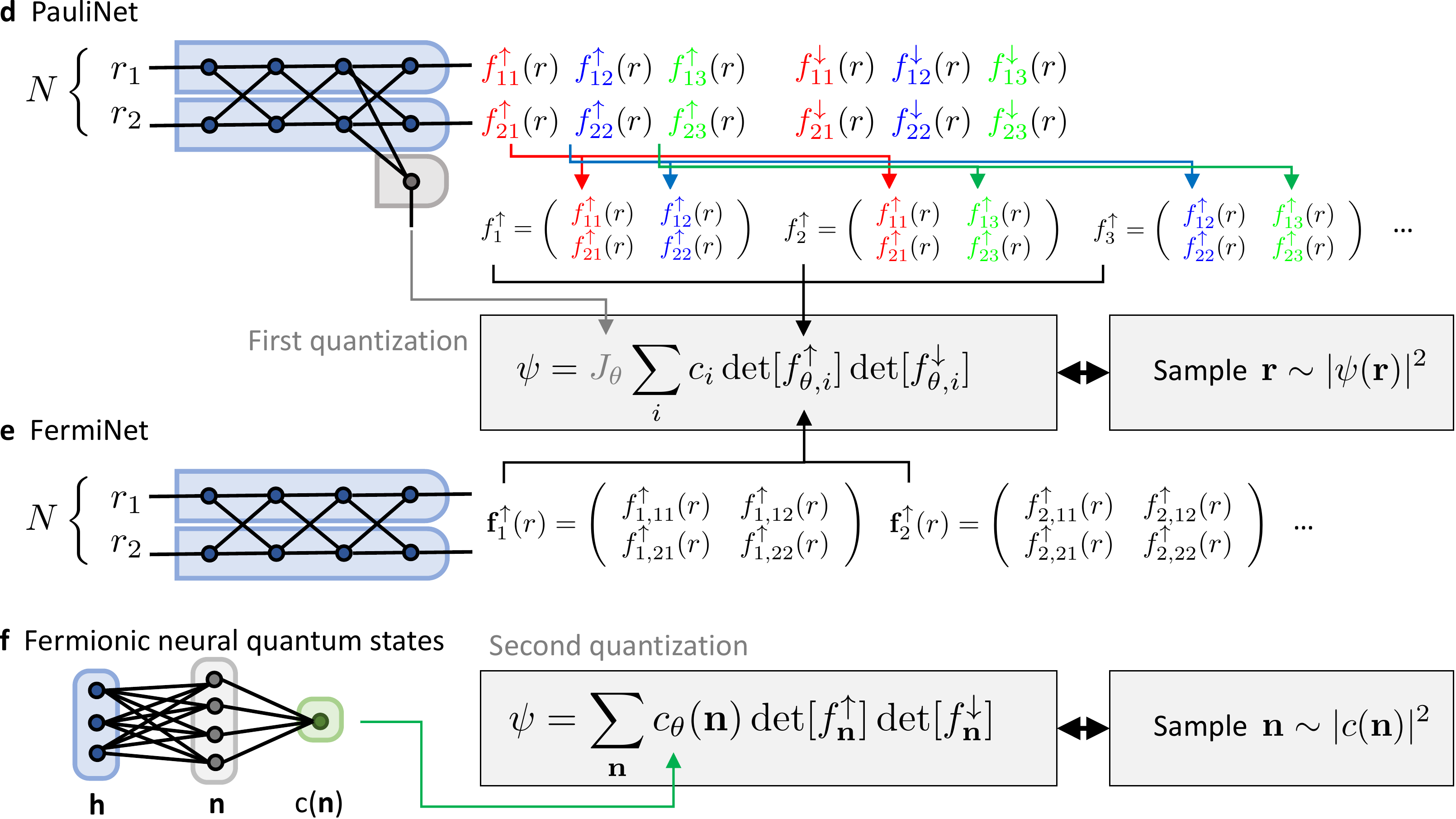}
\caption{\textbf{Electronic structure problem and its neural-network solutions.}
(\textbf a) The problem is fully specified by the geometry of a molecule and the electronic Schrödinger equation.
(\textbf b) Only fully antisymmetric wavefunctions are admissible as solutions due to the Pauli exclusion principle and (\textbf c) these are often represented with Slater determinants.
(\textbf d,\textbf e) Solutions formulated in first quantization use antisymmetric neural networks to represent the wavefunction directly in real space.
(\textbf f) Second quantization transfers the antisymmetry to a fixed finite basis, enabling the use of vanilla neural networks.
}\label{fig:intro}
\end{figure*}

\Cref{sec:el-structure} briefly reviews the components of electronic structure theory necessary for the development of the ML methods to be discussed later on.
The electronic structure problem is mapped to ML in \cref{sec:ml-for-se}, which is followed by a review of the ab-initio ML methods for QC formulated in real space and in a discrete basis in \cref{sec:real-space,sec:discrete-space}, respectively.
The review is concluded in \cref{sec:conclusion}.

\section{Electronic structure}
\label{sec:el-structure}

\subsection{Schrödinger equation}

QC aims at finding approximate solutions of the electronic Schrödinger equation that strike a good balance between accuracy and efficiency \citep{Piela20} (\cref{fig:QC}b).
The non-relativistic electronic Schrödinger equation within the Born--Oppenheimer approximation for a given molecule specified by the charges and coordinates of the nuclei, $Z_I$, $\mathbf R_I$, is a second-order differential equation for the wavefunction, $\psi(\mathbf r_1,\ldots,\mathbf r_N)$, which is a function of the coordinates of $N$ electrons (\cref{fig:intro}a):
\begin{gather}
  \hat H\psi(\mathbf r_1,\ldots,\mathbf r_N)=E\psi(\mathbf r_1,\ldots,\mathbf r_N), \label{eq:schrodinger} \\
\hat{H}:=\sum_{i}\bigg(-\tfrac{1}{2}{\nabla}_{\mathbf{r}_{i}}^{2}-\sum_{I}\frac{Z_{I}}{|\mathbf{r}_{i}-\mathbf{R}_{I}|}\bigg)+\sum_{i<j}\frac{1}{|\mathbf{r}_{i}-\mathbf{r}_{j}|}.
\label{eq:hamiltonian}
\end{gather}

An alternative formulation of the Schrödinger equation uses the notion of an \emph{expectation value},
\begin{equation}
  \begin{aligned}
  \langle\hat H\rangle_\psi
  &\equiv\frac{\langle\psi|\hat H|\psi\rangle}{\langle\psi|\psi\rangle} \\
  &=\frac{\int\mathrm d\mathbf r_1\cdots\mathrm d\mathbf r_N\psi(\mathbf r_1,\ldots,\mathbf r_N)\hat H\psi(\mathbf r_1,\ldots,\mathbf r_N)}{\int\mathrm d\mathbf r_1\cdots\mathrm d\mathbf r_N\lvert\psi(\mathbf r_1,\ldots,\mathbf r_N)\rvert^2}.
  \end{aligned}\label{eq:observable}
\end{equation}
Instead of solving \cref{eq:schrodinger}, the ground-state (lowest-energy) solution can be found by minimizing this energy expectation value with respect to all possible wavefunctions (variational principle),
\begin{equation}
  \label{eq:variational-principle}
  E=\min_{\psi}\langle\hat H\rangle_\psi .
\end{equation}

\subsection{Antisymmetric wavefunctions}

Electrons are fermions, and as such their wavefunction must be antisymmetric with respect to exchange of any two electrons.
This cardinal feature of electronic wavefunctions permeates the whole of QC\@.
In general, electrons also possess spin coordinates, $s_i\in\left\{\uparrow,\downarrow\right\}$, but the nonrelativistic Hamiltonian does not operate on spin, so the spin coordinate of each electron can be considered fixed.
To simplify the presentation here \parencite[for full treatment, see][Sec.~IV.E]{FoulkesRMP01}, we take advantage of the fixed spin coordinates, so the \emph{spatial} wavefunction must be antisymmetric only with respect to the exchange of same-spin electrons, i.e., when $s_i=s_j$ (\cref{fig:intro}b),
\begin{equation}
  \psi(\ldots,\mathbf r_i,\ldots,\mathbf r_j,\ldots)=-\psi(\ldots,\mathbf r_j,\ldots,\mathbf r_i,\ldots) .
  \label{eq:antisymmetry}
\end{equation}

By far the most common way to form antisymmetric wavefunctions in QC is as antisymmetrized products of single-electron functions (orbitals), $\phi_j(\mathbf r)$.
These products can be written as determinants of an $N\times N$ matrix, $\phi_j(\mathbf r_i)$, formed by putting $N$ electrons into $N$ orbitals, and are referred to as Slater determinants (\cref{fig:intro}c):
\begin{equation}
\label{eq:determinant}
D_{\boldsymbol\phi}(\mathbf{r}_1,\ldots,\mathbf{r}_N) = \frac{1}{\sqrt{N!}} \begin{vmatrix}
\phi_1(\mathbf{r}_1) & \phi_1(\mathbf{r}_2) & \cdots & \phi_1(\mathbf{r}_N) \\
\phi_2(\mathbf{r}_1) & \phi_2(\mathbf{r}_2) & \cdots & \phi_2(\mathbf{r}_N) \\
\vdots      & \vdots      & \ddots & \vdots      \\
\phi_N(\mathbf{r}_1) & \phi_N(\mathbf{r}_2) & \cdots & \phi_N(\mathbf{r}_N)
\end{vmatrix} .
\end{equation}
When interpreting $\phi_j(\mathbf r_i)$ as the $j$-the component of a $N$-dimensional feature vector for the $i$-th electron (using ML parlance), $\boldsymbol\phi(\mathbf r_i)$, a Slater determinant is in fact the only antisymmetric function of $N$ feature vectors that is linear in every one of them, making it a natural choice.
Alternative antisymmetric forms exist, such as the Pfaffian \citep{BajdichPRL06} or the Vandermonde determinant and its generalizations \citep{HanJCP19,AcevedoDLS20}, but these are far less common and we will not discuss them here.

Slater determinants formed from different orbitals can be further mixed in a linear combination without breaking the antisymmetry (\cref{fig:intro}c).
In fact, this simple technique is the powerhouse behind all the high-accuracy methods of QC, yet it is also its bane, because the number of Slater determinants required to achieve a given accuracy rises exponentially with the number of atoms in most cases.
For fermionic wavefunctions there is no known general approach to effectively reduce the search space from this exponential regime without sacrificing accuracy.
However, QC has produced many methods that achieve excellent approximations for specific molecules and materials of practical interest.
The cost of these highly accurate methods is generally less than exponential, but nevertheless increases rapidly with system size (\cref{fig:QC}b).

\subsection{Variational wavefunction methods}

An important class of QC methods derives directly from the variational principle (\pcref{eq:variational-principle}), by assuming a certain wavefunction \emph{ansatz}, $\psi(\cdot;\boldsymbol\theta)$, parametrized by $\boldsymbol\theta$.
Minimizing the energy of this ansatz with respect to $\boldsymbol\theta$ then always yields an upper bound for the exact ground-state energy,
\begin{equation}
  E=\min_{\psi}\langle\hat H\rangle_\psi\leq\min_{\boldsymbol\theta}\langle\hat H\rangle_{\psi(\cdot;\boldsymbol\theta)} .
\end{equation}
The bound becomes tighter as the expressiveness of the ansatz is improved.

One can distinguish two strategies to construct the ansatzes.
First, traditional QC uses relatively simple forms, such that the integral of \cref{eq:observable} can be evaluated analytically, which drastically simplifies the minimization problem \citep{Szabo96,Piela20}.
Second, quantum Monte Carlo (QMC) enables the use of arbitrarily complex ansatzes at the cost of having to do the integral evaluation and minimization stochastically \citep{Becca17}.
The latter is a natural framework to incorporate neural networks, and we introduce it in more detail in \cref{sec:qm-ml}.

Here we introduce three ansatzes for electronic wavefunctions of the first (traditional) kind, since they serve as scaffolding for the neural-network ansatzes of \cref{sec:real-space,sec:discrete-space}.
We also briefly discuss how they relate to other popular QC methods.

\begin{mybox}[label=box:first-second-quant]{First and second quantization}

\noindent
Computational methods for the electronic Schrödinger equation can be divided to first-quantized approaches in real space and second-quantized approaches in a discrete basis.
In first quantization, one works with the individual electrons and their coordinates directly in real space ($\mathbf r_i\in\mathbb R^3$, $i=1,\ldots,N$) as in \cref{eq:schrodinger},
\begin{equation*}
\lvert\psi\rangle
=\int\mathrm d\mathbf r_1\mathrm d\mathbf r_2\!\cdots\psi(\mathbf r_1,\mathbf r_2,\ldots)\lvert\mathbf r_1\mathbf r_2\!\cdots\rangle, \\
\end{equation*}
Here, $\psi$ must be an antisymmetric function, which specifies which electrons occupy which coordinates, while the many-electron basis states ($\lvert\mathbf r_1\mathbf r_2\cdots\rangle$) are ordinary non-symmetric (Cartesian) product states.

In second quantization, one has to first introduce a discrete basis (in practice finite), labelled by $k$, which then enables one to work with preformed antisymmetric many-electron basis states (Slater determinants), and rather than specifying which electrons occupy which one-electron states, the occupation numbers ($n_k\in\{0,1\}$, $\sum_k n_k=N$) specify which one-electron states are occupied without any reference to a particular electron,
\begin{equation*}
\lvert\psi\rangle=\sum_{n_1n_2\cdots}\psi_{n_1n_2\cdots}\lvert n_1n_2\!\cdots\rangle .
\end{equation*}
Here, $\psi_{n_1n_2\cdots}$ can be an arbitrary tensor without antisymmetry, which is instead encoded in the many-electron basis states $\lvert n_1n_2\cdots\rangle$.

This ability to push the antisymmetry from the wavefunction object to the many-electron basis is the main advantage of second quantization, at the cost of having to commit to a particular discrete basis.
But regardless of the computational framework, either the wavefunction object itself (in first quantization) or the many-electron basis (in second quantization) consists of Slater determinants, and in high-accuracy methods their number grows rapidly with system size.
{\vspace{0.5em}\begin{center}
\includegraphics[width=6.5cm]{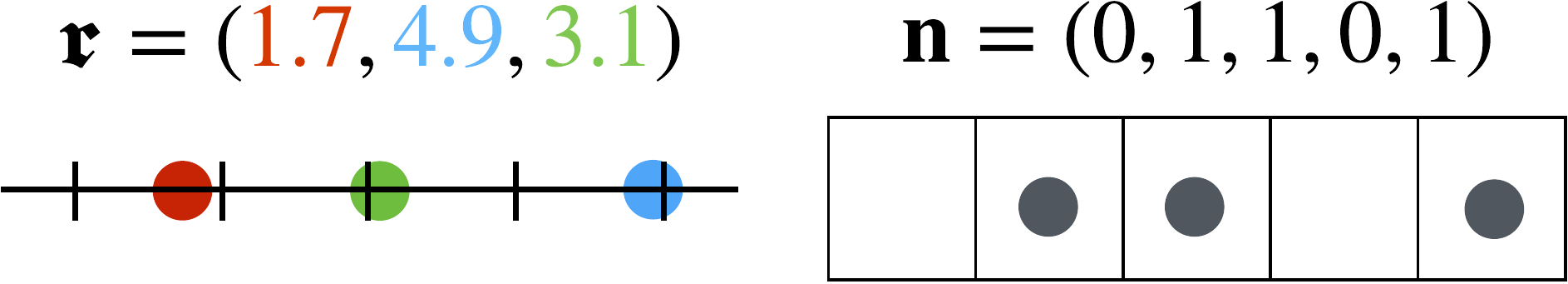}
\end{center}
{\small\textbf{First and second quantization.} Illustration on $N=3$ electrons in 1D and a finite basis of size 5.}
$\bm{\mathfrak{r}}=(\mathbf{r}_1, \mathbf{r}_2, \mathbf r_3)$.
}
\end{mybox}

\paragraph{Hartree--Fock}

Perhaps the simplest nontrivial ansatz in QC is the single Slater determinant of \cref{eq:determinant}, where the orbitals $\phi_j(\mathbf r)$ are considered as free parameters.
Optimized variationally, this ansatz leads to the so-called Hartree--Fock (HF) method.
In practice the orbitals are linearly expanded in a fixed finite one-electron basis, $\varphi_k(\mathbf r)$, $k=1,\ldots,K$, with $K \sim N$ in most cases:
\begin{equation}
E_\text{HF}=\min_{\phi_j}E[\det\phi_j(\mathbf r_i)]
\approx\min_{C_{kj}} E\big[\textstyle\det\sum_k C_{kj}\varphi_k(\mathbf r_i)\big] .
\label{eq:hartree-fock}
\end{equation}
The use of a finite basis set turns the functional optimization problem of \cref{eq:hartree-fock} into a computational problem whose cost scales with the fourth power of the number of basis functions, $O(K^4)$, assuming a naive implementation.
On its own, the HF ansatz is expressive enough to describe much of chemistry qualitatively, but not always and certainly not quantitatively.
However, it can be considered a starting point for most wavefunction-based QC methods.

Density functional theory (DFT) is not such a method, relying instead on an in-principle exact mapping of the ab-initio Hamiltonian (\pcref{eq:hamiltonian}) to a mean-field-like problem, which can be solved exactly with a single Slater determinant \citep{JonesDFTReview20215,TealePCCP22}.
However, the variational principle does not hold in DFT because the exchange-correlation contributions to the energy functional are not known exactly and must be approximated in practice.
From here on, we will stay within the variational principle and instead focus on increasing the expressiveness of the HF ansatz.

\paragraph{Configuration interaction}

The HF ansatz can be straightforwardly extended by forming multiple Slater determinants from different sets of orbitals and considering their linear combination (\cref{fig:intro}c),
\begin{equation}
  \psi(\mathbf r_1,\ldots,\mathbf r_N)=\sum_p c_p D_{\boldsymbol\phi_p}(\mathbf r_1,\ldots,\mathbf r_N) .
  \label{eq:ci}
\end{equation}
When the orbitals of each determinant are pooled from a larger superset of (mutually orthogonal) \emph{fixed} orbitals of size $M>N$, and the only free parameters are the linear coefficients of the determinants, the ansatz is called \emph{configuration interaction} (CI).
One of the appeals of the CI ansatz is that its Slater determinants can be considered a many-electron antisymmetric basis and labelled using the occupation numbers of the one-electron states.
This so-called \emph{second quantized} formalism has many convenient properties for computation (see Box~\ref{box:first-second-quant}).
The simplest version of CI, called full CI (FCI), considers all $\binom MN$ possible Slater determinants and is exact within the chosen finite one-electron basis.
In the usual case when $M \sim N$, however, the computational effort scales exponentially with $N$, which makes FCI applicable only to the smallest molecules.
Ways to tackle the exponential scaling include fixed truncation of the CI expansion or its ``compression'' through analytical means (coupled cluster theory, \cite{Bartlett2007}; matrix product states, \cite{ChanDMRG2011}), deterministic pruning (selected CI, \cite{HuronJCP73}), or stochastic sampling (FCI-QMC, \cite{BoothJCP09}). 
\Cref{sec:discrete-space} explores a novel way of ``compressing'' the CI expansion through neural networks.

\paragraph{Beyond fixed bases}

The effectiveness of the CI ansatz depends on the choice of the fixed molecular orbitals $\phi_j(\mathbf{r})$ from which the Slater determinants $D_{\boldsymbol\phi_p}(\mathbf{r}_1,\ldots,\mathbf{r}_N)$ are built.
A natural extension of CI allows both the orbitals and the CI expansion coefficients $c_p$ to vary during the variational minimization.
Such an ansatz of two stacked linear combinations (\pcref{eq:hartree-fock,eq:ci}) is harder to optimize but much more expressive.
The most common variant is to consider all $\binom{M'}{N'}$ Slater determinants formed by letting $N'<N$ electrons occupy a space of $M'<M$ orbitals, while the remaining $N-N'$ electrons occupy a fixed set of inactive obitals.
This is called the \emph{complete active space self-consistent field} (CASSCF) method \citep{OlsenIJQC11}.
Due to the larger variational freedom, a CASSCF ansatz typically requires many fewer determinants than a CI ansatz of comparable accuracy.

But CASSCF and even FCI are still limited by the fixed one-electron basis used to form the molecular orbitals (\pcref{eq:hartree-fock}): FCI is only exact in the complete basis set limit, which in practice cannot be reached for any but the smallest molecular systems.
An extension of the CASSCF ansatz would allow not only the one-electron orbitals but also the one-electron basis functions to vary.
The stacked structure of such an ansatz would be reminiscent of deep neural networks, and \Cref{sec:real-space} explores the culmination of this line of thought by incorporating actual deep neural networks into the ansatz.
This removes any \emph{a priori} limitations on the expressiveness.
By making each individual determinant maximally expressive, such ansatzes further reduce the number of determinants required to reach a given accuracy.

\section{Machine learning for electronic Schrödinger equation}
\label{sec:ml-for-se}

\begin{mybox}[label=sec:QMC]{Variational Monte Carlo}

\noindent
Optimization of wavefunctions with neural networks naturally leads to the variational Monte Carlo (VMC) framework.
First, Monte Carlo integration of \cref{eq:observable} can handle arbitrarily complicated ansatzes for which analytical integrals are not available.
Second, VMC samples these integrals stochastically which naturally combines with the stochastic gradient descent used for optimizing neural networks.
In traditional QC, VMC has been used extensively with real-space first-quantized approaches \citep{FoulkesRMP01} and more recently in the discrete-basis second-quantized setting \citep{NeuscammanJAGPHilbert2013, SabzevariJCTC18}.

The expectation value of any operator, such as the Hamiltonian (\pcref{eq:observable}), can be written as a Monte Carlo integral over a continuous or discrete basis, $\{\lvert\mathbf x\rangle\}$,
\begin{equation*}
\langle \hat H \rangle_\psi
=\int_\mathbf x\frac{\lvert\langle\mathbf x|\psi\rangle\rvert^2}{\langle\psi|\psi\rangle}\frac{\langle\mathbf x|\hat H|\psi\rangle}{\langle\mathbf x|\psi\rangle}
=\mathbb E_{\mathbf x\sim\lvert\langle\mathbf x|\psi\rangle\rvert^2}\left[E_\text{loc}(\mathbf x)\right] .
\end{equation*}
Here, the expectation value is obtained as an expected value of a ``local'' energy $E_\text{loc}$, local in the sense that it is defined for every basis element $\mathbf x$.

A straightforward and generally applicable way to obtain the samples is Markov-chain Monte Carlo (MCMC).
MCMC is an iterative procedure, in which a new sample point, $\mathbf x'$, is produced from a current one, $\mathbf x$, by making a proposal step with probability $g(\mathbf x'|\mathbf x)$, and then accepting or rejecting the proposal with probability
\begin{equation*}
p=\min\left(1,\frac{\lvert\langle\mathbf x'|\psi\rangle\rvert^2g(\mathbf x|\mathbf x')}{\lvert\langle\mathbf x|\psi\rangle\rvert^2g(\mathbf x'|\mathbf x)}\right) .
\end{equation*}
The resulting Markov chain then samples $\lvert\langle\mathbf x|\psi\rangle\rvert^2$.
Variants of MCMC differ in the construction of the proposal steps and $g$, and include the simplest Metropolis algorithm ($g(\mathbf x'|\mathbf x) = g(\mathbf x|\mathbf x')$) as well as more sophisticated flavours such as Langevin Monte Carlo.

The VMC formula for the expectation value is exact in the limit of infinite sample size, $\mathcal N\rightarrow\infty$, but in practice it incurs a statistical error proportional to $\sqrt{\mathrm{Var}[E_{\textrm{loc}}] / \mathcal N}$.
While $1/\sqrt{\mathcal N}$ converges slowly with sample size, VMC has the great benefit that the as the ansatz converges to the exact eigenstates, the local energy converges to a constant (the exact energy), and as such its variance vanishes and so does the statistical sampling error.
\end{mybox}

\subsection{Mapping quantum mechanics to machine learning}
\label{sec:qm-ml}

\begin{figure}[t]
\centering
\includegraphics{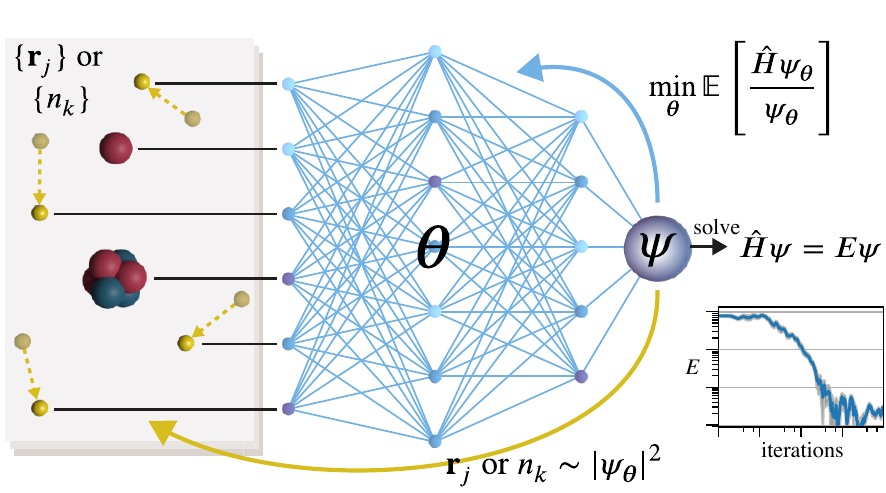}
\caption{\textbf{Variational Monte Carlo with neural networks.}
Electron positions, $\mathbf r_j$, or orbital occupation numbers, $n_k$, describe an electron configuration which is an input to the wavefunction, $\psi$, represented by a neural network parametrized with $\boldsymbol\theta$.
The wavefunction is used in two ways: first, to sample new electron configurations which provide new input to the neural network (yellow), and second, to evaluate the electronic energy, which is minimized by varying the network parameters (blue).
}\label{fig:vmc-nn}
\end{figure}

\begin{table}[t]
\centering
\caption{\textbf{Dictionary of electronic structure and machine learning.}
}\label{tab:mapping}
\begin{tabular}{cc}
    \toprule
    Electronic structure & Machine learning \\
    \midrule
    Wavefunction & Probability distribution \\
    Natural orbital & Marginal distribution \\
    Stochastic reconfiguration & Natural gradient descent \\
    Hartree--Fock & Mean-field variational Bayes \\
    Diffusion Monte Carlo & Particle filtering;  \\
     & Sequential Monte Carlo \\
    \bottomrule
\end{tabular}
\end{table}

A ML problem and its solution are specified by the model, its inputs and outputs, the data, and the optimization criterion (loss function).
In this regard, solving the Schrödinger equation with the variational principle amounts to the following ML problem (\cref{fig:vmc-nn}).
The neural network (\cref{sec:deep-learning}) represents a wavefunction, which accepts electron coordinates (first quantization) or occupation numbers (second quantization) as input and outputs the wavefunction value.
The loss function is the energy expectation value corresponding to this wavefunction.
The inputs are sampled from the probability distribution given by the square of the wavefunction represented by the current neural network, and the Hamiltonian operator is used to obtain an estimate of the loss function from the samples.
The parameters of the network, and thus the wavefunction, are then modified to minimize the loss function.
Except for the representation of the wavefunction as a network, this is the regular variational Monte Carlo (VMC) framework (Box~\ref{sec:QMC}).
The optimization methods used (Box~\ref{sec:optimization}) are also fairly conventional, although adapted to a neural network context.
This straightforward correspondence between the Schrödinger equation and ML led to the introduction of similar concepts on both sides, albeit known under different names (\cref{tab:mapping}).

The applicability of deep learning for quantum-mechanical calculations was first realized and exploited by \citet{CarleoS17} for the case of spin lattices in one and two dimensions.
Their approach, known as Neural Quantum States (NQS), has since been applied to many different quantum systems \citep{saito2017solving, nomura2017restricted, corey2021variational, nikita2021broken}.
In essence, this review is concerned with the extension of this approach to electrons in molecules.

\begin{mybox}[label=sec:optimization]{Optimizing neural-network ansatzes}

\noindent
Up to the statistical error, the VMC expectation value for the energy (Box~\ref{sec:QMC}) obeys the variational principle (\pcref{eq:variational-principle}).
VMC exploits this by varying a parametric wavefunction ansatz $\psi_{\boldsymbol{\theta}}$ so as to minimize the energy.
For a sufficiently expressive ansatz, the variational energy will eventually approximate the ground state energy of \cref{eq:schrodinger} and the ansatz will approximate the ground state wavefunction $\Psi$.

The most straightforward optimization method is gradient descent, where the parameters are iteratively updated as
\begin{equation*}
    \boldsymbol{\theta} \leftarrow  \boldsymbol{\theta} - \eta\frac{\partial E(\boldsymbol{\theta})}{\partial\boldsymbol{\theta}}
\end{equation*}
with learning rate $\eta > 0$.
The energy gradient is given by
\begin{equation*}
     \frac{\partial E}{\partial \theta_k}  = \langle \hat O^{*}_k \hat H \rangle - \langle \hat H \rangle \langle \hat O^{*}_k \rangle ,
\end{equation*}
where
\begin{equation*}
\hat O_{k}(\mathbf x)  = \frac{\partial\ln \psi(\mathbf x; {\boldsymbol{\theta}})}{\partial \theta_k}
\end{equation*}
is an operator representing the logarithmic derivatives of the wavefunction.
This gradient can be efficiently estimated using Monte Carlo integration (Box~\ref{sec:QMC}).

In some cases the optimization can be sped up and made more stable with higher-order methods, such as the stochastic reconfiguration (SR) scheme \citep{SorellaPRL98}.
SR takes the correlation between individual variational parameters into account by introducing the quantum geometric tensor $S$:
\begin{equation*}
    S_{kk'} = \langle O_{k}^{*}O_{k'} \rangle - \langle O_{k}^{*}\rangle\langle O_{k'}\rangle .
\end{equation*}
The update rule is then modified to
\begin{equation*}
    \boldsymbol{\theta} \leftarrow  \boldsymbol{\theta} - \eta \mathbf S^{-1} \frac{\partial E(\boldsymbol{\theta})}{\partial\boldsymbol{\theta}} .
\end{equation*}
The SR scheme approximates an imaginary-time evolution where each iteration tries to best approximate the state $\mathrm e^{-\eta \hat H}\ket{\psi}$.
SR is similar to the natural gradient descent algorithm \citep{amari_natural_1998} that is well-known in the ML community, and $\mathbf S$ can be interpreted as a quantum generalization of the Fisher information matrix \citep{Ay17}.
In some cases, it is convenient to approximate the quantum geometric tensor $\mathbf S$ using the Kronecker-factored approximate curvature (KFAC) approach \citep{martens2015optimizing}.
\end{mybox}

\subsection{Deep learning}
\label{sec:deep-learning}

The standard practice in ab-initio QC today is in some ways analogous to the state of computer vision before the rise of deep learning.
Prior to 2012, the best pipelines for large-scale image recognition consisted of a combination of hand-designed features and simple ML models \citep{perronnin2010large}.
A single deep convolutional neural network trained end-to-end was able to cut the recognition error in half relative to these systems \citep{krizhevsky2012imagenet}, and since then deep neural networks have dominated computer vision research.

In ab-initio QC, ground-state solutions to the Schrödinger equation are usually represented by a wavefunction ansatz with a relatively simple functional form, and parameters are usually fit through a mix of procedures (fixed-point iteration, variational optimization) rather than a unified end-to-end estimation of all parameters simultaneously.
The development of deep QMC methods is driven by the hope that the use of neural networks will significantly increase the expressiveness of wavefunction ansatzes, enabling large leaps in accuracy as in image recognition.
To appreciate how and why deep neural networks can be usefully applied in QC, a brief review of their application in artificial intelligence is necessary.
For a thorough review of the history of deep learning, see \citet{schmidhuber2015deep}, and for a review of the fundamental concepts in deep learning, see \citet{lecun2015deep}.

Neural networks date back to the very beginning of computer science \citep{mcculloch1943logical}, and their modern form originates with the single perceptron ``unit'' \citep{rosenblatt1958perceptron}, which produces as output a non-linear function of the sum of a constant, known as the bias, and a linear combination of its inputs.
The non-linear function rises from zero to one as its input increases, mimicking the activation function of a biological neuron.
When many such units are assembled in parallel to form a ``layer,'' and several layers are computed serially, taking the output from one layer as the input to the next, the resulting multi-layer perceptron (MLP) can, in theory, represent any smooth function to arbitrary accuracy given enough units \citep{hornik1989multilayer}.
However, actually fitting or \emph{learning} a set of parameters that matches any given function is different matter.
A form of gradient descent utilizing derivatives computed using backpropagation, or reverse-mode automatic differentiation \citep{werbos1974beyond,linnainmaa1970representation, linnainmaa1976taylor}, was found to be effective for training neural networks \citep{rumelhart1986learning}.
This led to a wave of enthusiasm for neural networks, which eventually faded as several issues were discovered, such as the infamous ``vanishing gradients'' and getting stuck in local minima.

Several factors were instrumental in rehabilitating neural networks under the banner of ``deep learning'': a combination of algorithmic advances \citep{glorot2010understanding} and the use of modern GPU hardware \citep{hooker2020hardware} made the computations much faster, and the resulting ability to train larger networks made issues with local minima less severe \citep{dauphin2014identifying, choromanska2015loss}.
Furthermore, deep neural networks with the help of stochastic gradient descent can be applied straightforwardly and efficiently to large datasets, unlike other ML models \citep{bottou2008learning, bottou2011tradeoffs}.
Finally, empirical successes like winning the ImageNet Large Scale Visual Recognition Challenge \citep{russakovsky2015imagenet} helped legitimize deep learning research and generate excitement among researchers.

Today, the barrier to entry for developing and training deep neural networks is quite low, thanks to a mature ecosystem of software libraries for numerical computing with automatic differentiation and hardware accelerators \citep{AbadiOSDI16, paszke2017automatic, bradbury2018jax}.
However, actually achieving good performance from a deep learning model still requires some finesse and application of various heuristics.
It is safe to say that a significant amount of the practice of deep learning remains more art than science.
The good news is that once effective heuristics for a particular problem domain have been developed, these same heuristics can often be applied with little modification to other problems in the same domain.

\subsection{Neural network architectures}

The starting point for most neural networks is the multi-layer perceptron (MLP), formed as a composition of $L$ layers,
\begin{equation}
\begin{aligned}
\mathrm{MLP}(\mathbf{x}) &= f^L \circ f^{L-1} \circ \cdots \circ f^{1} (\mathbf{x}) , \\
f^{\ell}(\mathbf{z}) &= f\left(\mathbf{W}^{\ell} \mathbf{z} + \mathbf{b}^\ell\right) ,
    \label{eqn:mlp}
\end{aligned}
\end{equation}
where $f$ is some non-linear \emph{activation} function, and $\mathbf{W}^\ell$ and $\mathbf{b}^\ell$ are the matrices of weights and vectors of biases to learn.
While a vanilla MLP is capable of representing arbitrary functions, the real power of neural networks comes from more sophisticated architectures.
Many of these architectures are designed to encode some particular \emph{invariance} or \emph{equivariance}---that is, when the input to the network is transformed in a particular way, the output should either be unchanged or should transform in a corresponding way.
For instance, the weights in a layer of a convolutional neural network (ConvNet) \citep{lecun1998gradient} are restricted to be a discrete convolution operator,
which constrains each layer to be translation-equivariant, a natural constraint for image recognition, and also dramatically reduces the number of possible weights in a layer.

Equivariance to permutation is another frequently useful property, and one that is especially important in real-space approaches to representing electronic wavefunctions (see \cref{sec:real-space}).
A simple permutation-equivariant layer first proposed by \citet{shawe1989building} can be constructed by applying the same transformation to each input and summing the results.
More sophisticated permutation-equivariant layers are used by models like the Transformer \citep{vaswani2017attention} or SchNet \citep{schutt2018schnet}.
Many of these equivariant layers can be unified in a conceptual framework based around the language of geometry and group theory, wherein the choice of transformation to be equivariant to leads naturally to recipes for constructing the appropriate neural network layers \citep{bronstein2021geometric}.

Another class of neural network architectures, which have been influential as wavefunction ansatzes, are restricted Boltzmann machines (RBMs) \citep{hinton2006reducing}.
These were originally developed for unsupervised learning, but in the VMC setting considered here they lead to a simple deterministic expression for the log probability that closely resembles a one-layer MLP\@.
Despite their early popularity, RBMs have been largely eclipsed in the AI community by other methods for unsupervised learning, such as variational autoencoders \citep{kingma2013auto}, generative adversarial networks \citep{goodfellow2014generative}, normalizing flows \citep{rezende2015variational}, autoregressive models \citep{oord2016wavenet, oord2016conditional}, and diffusion models \citep{sohl2015deep}.
In fact, some of these newer models have started to have an impact as neural network wavefunction ansatzes for spin systems.
Examples are deep autoregressive quantum states \citep{sharir2020deep}, convolutional neural networks \citep{choo2019two}, recurrent neural networks \citep{hibat-allah_recurrent_2020}, and normalizing flows \citep{xie_ab-initio_2021}.

\section{Electrons in first quantization}
\label{sec:real-space}

\begin{figure*}[t]
    \centering
    \includegraphics[width=\textwidth]{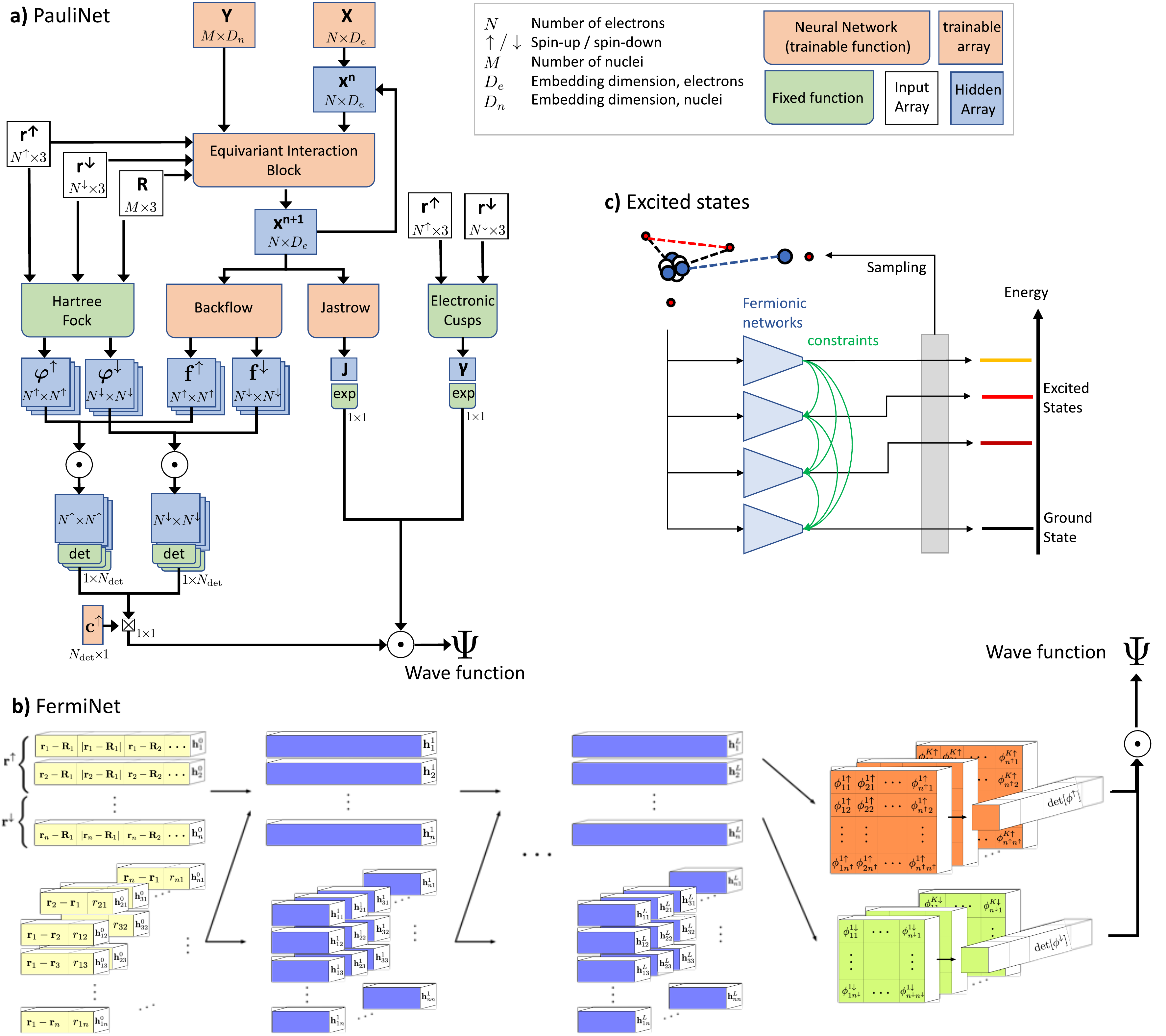}
    \caption{\textbf{Neural-network architectures for selected real-space wavefunctions}.
    (\textbf a) Original PauliNet architecture from \citep{HermannNC20}.
    (\textbf b) Original FermiNet architecture from \citep{PfauPRR20}.
    Both architectures have been modified and extended by various contributions mentioned in this review.
    (\textbf c) Approach to computing exciting states in \citep{Entwistle22}.
    }\label{fig:real_space_networks}
\end{figure*}

One approach to studying the electronic problem with deep learning is to work with parameterized many-body wavefunctions in first quantization, $\psi(\bm{\mathfrak{r}}; {\boldsymbol\theta})$.
Here  $\bm{\mathfrak{r}}$ stands for the $N$-tuple of electron coordinates, $\mathbf{r}_1,\mathbf{r}_2,\ldots,\mathbf{r}_N$, and sampling is realized over electronic positions $\bm{\mathfrak{r}}$ (Box~\ref{sec:QMC}).
The antisymmetry constraint (\pcref{eq:antisymmetry}) must be imposed in $\psi$ to avoid collapsing onto a lower-energy bosonic state.
A commonly adopted form is $\psi(\bm{\mathfrak{r}}; {\boldsymbol\theta})=S(\bm{\mathfrak{r}}; {\boldsymbol\theta})\times A(\bm{\mathfrak{r}}; {\boldsymbol\theta})$, where the first factor is symmetric (or ``bosonic'') under exchange of electron coordinates and the second factor carries the necessary antisymmetry.
The simplest and most common approach is to build the antisymmetric part of the wavefunctions using Slater determinants (\pcref{eq:determinant}).
As discussed in \cref{sec:el-structure}, single Slater determinants with fixed orbitals have limited expressiveness and many such determinants need to be combined to achieve high accuracy.
A natural generalization of a sum of fixed-orbital Slater determinants is the commonly-used Slater--Jastrow wavefunction
\begin{equation}
\psi(\bm{\mathfrak{r}}; \boldsymbol\theta) = \mathrm e^{J\left(\{\bm{\mathfrak r}\};\boldsymbol\theta\right)} \sum_k c_k \begin{vmatrix}
\phi_{1}^k(\mathbf{r}_1;\boldsymbol\theta)  & \cdots & \phi_{1}^k(\mathbf{r}_N;\boldsymbol\theta) \\
\vdots     & \ddots & \vdots      \\
\phi_{N}^k(\mathbf{r}_1;\boldsymbol\theta)  & \cdots & \phi_{N}^k(\mathbf{r}_N;\boldsymbol\theta)
\end{vmatrix}
\label{eq:SJ}
\end{equation}
where the Jastrow factor, $J(\{\bm{\mathfrak{r}}\}; {\boldsymbol\theta})$ constitutes the symmetric (``bosonic'') part of the state and typically contains one- and two-body (and in many cases higher-order) parameterized correlations.
The set notation, $\{\bm{\mathfrak{r}}\} \equiv \{\mathbf{r}_1,\ldots,\mathbf{r}_2\}$, indicates that $J$ does not depend on the order of the electron coordinates.
The determinants in \cref{eq:SJ} are typically replaced with the product of spin-up and spin-down determinants \citep{FoulkesRMP01}.
Separating the up- and down-spin determinants improves computational efficiency, simplifies the implementation, and makes it easier to handle the electron-electron cusps, while leaving expectation values of spin-independent operators unchanged.

More flexible parametric forms can be obtained leveraging the approximation power of artificial neural networks.
In the following, we discuss neural-network-based strategies to parameterize these forms.

\subsection{Discrete space}

The first applications of neural networks to electronic systems were for electrons moving in discretized space, as realized, for example, in the 2D Hubbard model of strongly-interacting electrons.
In the following, for simplicity, we discuss the case of $N$ spinless electrons in $M$ lattice sites, and denote with $l(\mathbf{r}) \in [1,M]$ the discrete lattice index corresponding to electron position $\mathbf{r}$.
The extension to the spinful case will be considered more in detail when discussing continuous space later on.
The symmetric part $S(\bm{\mathfrak{r}}; {\boldsymbol\theta})$ can be readily parameterized with a strategy closely related to NQS for spins:

\begin{equation}
S(\bm{\mathfrak{r}}; {\boldsymbol\theta})=g(n(\bm{\mathfrak{r}}); {\boldsymbol\theta}),
\end{equation}
where $n(\bm{\mathfrak{r}})$ is the unique occupation number representation corresponding to the electronic positions $\bm{\mathfrak{r}}$ and $g$ represent a generic function which could be represented by a neural network.
Since the occupation numbers $n(\bm{\mathfrak{r}})$ are invariant under permutation of the electron positions, $g(n(\bm{\mathfrak{r}}))$ is also symmetric under exchange.
Any of the NN architectures also adopted for spin systems \citep{CarleoS17} or lattice bosons \citep{saito2017solving} can be used to represent the symmetric part $S$.
Early works on the Hubbard model adopted positive-definite RBM-based parameterizations of $S(\bm{\mathfrak{r}}; {\boldsymbol\theta})$ \citep{nomura2017restricted}, while more recent works have adopted deep-network parameterizations allowing for sign changes \citep{stokes_quantum_2020}.

The simplest parameterization for the antisymmetric part, $A(\bm{\mathfrak{r}}; {\boldsymbol\theta})$, is again a Slater determinant
\begin{equation}
\label{eq:atheta_discrete_space}
    A(\bm{\mathfrak{r}}; {\boldsymbol\theta}) = \begin{vmatrix}
\phi_{1}(\mathbf{r}_1; {\boldsymbol\theta})  & \cdots & \phi_{1}(\mathbf{r}_N; {\boldsymbol\theta}) \\
\vdots     & \ddots & \vdots      \\
\phi_{N}(\mathbf{r}_1; {\boldsymbol\theta})  & \cdots & \phi_{N}(\mathbf{r}_N; {\boldsymbol\theta})
\end{vmatrix},
\end{equation}
where the matrix $\Phi \in \mathbb{C}^{N \times M}$ of discrete orbitals $\phi_{i}(\mathbf{r}_j)=\Phi_{i,l(\mathbf{r}_j)}$ holds the variational parameters to be optimized.
This approach, however, has the important drawback of not providing enough variational flexibility, since it effectively fixes the anti-symmetric part to a mean-field reference solution.

\paragraph{Neural backflow}
A significant improvement is obtained by considering a many-body backflow transformation of the orbitals \citep{feynman_energy_1956,kwon_effects_1993}.
In this variational form, the matrix of one-electron orbitals $\Phi$ is promoted to a parameterized many-electron function depending on all the occupation numbers:

\begin{equation}
\Tilde\Phi_{ij}(\boldsymbol\theta)=\Phi_{ij}(\boldsymbol\theta) + \Delta_{ij}(n(\bm{\mathfrak{r}}) ; \boldsymbol\theta),
\end{equation}
where $\Delta$ is a correction to the single-particle orbitals $\Phi$.
In physics-inspired parameterizations, $\Delta$ is typically taken to be a simple function of the electronic occupation numbers \citep{tocchio_role_2008}.
The neural backflow method \citep{LuoPRL19} instead introduced a flexible parameterization of the backflow orbitals based on artificial neural networks.
In this case, $\Delta$ is parameterized with a MLP taking as inputs the electronic occupation numbers and outputing a many-body correction to the matrix $\Phi$.
This approach allows the orbitals to dynamically change depending on the positions of the electrons, thus allowing one to include genuinely many-body correlations in the antisymmetric part of the wavefunction.

\paragraph{Constrained hidden fermions}
Neural backflow transformations are not the only way to introduce flexible parameterizations of the antisymmetric part of the wavefunction.
The constrained hidden fermion formalism builds on the idea of introducing a set of $\tilde{N}$ auxiliary fermionic particles, with positions $\mathfrak{q}$, and living on $\tilde{M}$ lattice sites.
These auxiliary particles are used to effectively mediate correlations among the physical degrees of freedom \citep{robledo_moreno_fermionic_2022}.
Calling $\tilde{A}(\bm{\mathfrak{r}},\bm{\mathfrak{q}}; {\boldsymbol\theta})$ a Slater determinant for the extended (physical+hidden) system, the resulting antisymmetric form for the physical system is given by
\begin{equation}
   A(\bm{\mathfrak{r}}; \boldsymbol\theta)=\tilde{A}(\bm{\mathfrak{r}},F(\bm{\mathfrak{r}}; \boldsymbol\theta)).
\end{equation}
In this expression, $F$ is a function, parameterized by a neural network, mapping the physical positions to the hidden ones.
This approach has been shown to improve systematically over the neural backflow form for the 2D Hubbard model \citep{robledo_moreno_fermionic_2022}.

\subsection{Continuous space}

We now focus on describing the important case of first-quantized electrons in continuous space, directly corresponding to the electronic Schrödinger equation.
As in the discrete-space case, the Slater--Jastrow form may be improved in a matter suitable for use with neural quantum states by adding a backflow transformation, in which the one-electron orbitals $\phi_{i}(\mathbf{r}_j; {\boldsymbol\theta})$ are replaced by many-electron functions $\Tilde\phi_{i}(\mathbf{r}_j,\{\bm{\mathfrak{r}}\}; \boldsymbol\theta)$.
The backflow transformation can either modify the orbitals directly via a multiplicative and/or additive term:
\begin{gather}
\Tilde\phi_{i}(\mathbf{r}_j, \{\bm{\mathfrak{r}}\}; \boldsymbol\theta) = \phi_i(\mathbf{r}_j) f_{i}^{\bigotimes}(\mathbf{r}_j, \{\bm{\mathfrak{r}}\}; \boldsymbol\theta) +  f_{i}^{\bigoplus}(\mathbf{r}_j, \{\bm{\mathfrak{r}}\}; \boldsymbol\theta),
	\label{eq:backflow_mult_add}
\end{gather}
or act as a quasiparticle transformation of the electron coordinates:
\begin{equation}
    \Tilde\phi_{i}({\mathbf{r}_j}, \{\bm{\mathfrak{r}}\}; \boldsymbol\theta) = \phi_i\big(\mathbf{r}_j + \bm{\xi}(\{\bm{\mathfrak{r}}\}; {\boldsymbol\theta})\big),
    \label{eq:traditional-backflow}
\end{equation}
where the paramterized functions, $f_i^{\bigotimes}, f_i^{\bigoplus}, \bm{\xi}$, are invariant to permutations of $\{\bm{\mathfrak{r}}\}$, and $\bm{\xi}(\{\bm{\mathfrak{r}}\};\boldsymbol\theta)$ is a three-component vector that modifies $\mathbf{r}_j$.
If we consider a determinant of orbitals of this form,
\begin{equation}
	\begin{vmatrix}
		\phi_1(\mathbf r_1; \{\bm{\mathfrak{r}}\}) & \cdots & \phi_1(\mathbf r_N; \{\bm{\mathfrak{r}}\}) \\
		\vdots  & \ddots & \vdots \\
		\phi_N(\mathbf r_1; \{\bm{\mathfrak{r}}\}) &
		\cdots & \phi_N(\mathbf r_N; \{\bm{\mathfrak{r}}\}) \\
	\end{vmatrix},
	\label{eq:equivariant_det}
\end{equation}
then we see that orbitals with backflow transformations are just one example of a broader class of functions: in order for the determinant to be antisymmetric, the matrix with elements $\Phi_{ij} = \phi_i(\mathbf{r}_j; \{\mathfrak{r}\})$ must be permutation-equivariant; that is, exchanging electrons $k$ and $l$ also exchanges columns $k$ and $l$.
While traditional Slater--Jastrow--backflow wavefunctions have had considerable success, they also have limitations due to the choice of fixed functional forms.
The goal, therefore, is to come up with more flexible permutation-equivariant functions.
Here we highlight several approaches that share this common theme.

\begin{figure}[t]
    \centering
    \includegraphics[width=0.5\textwidth]{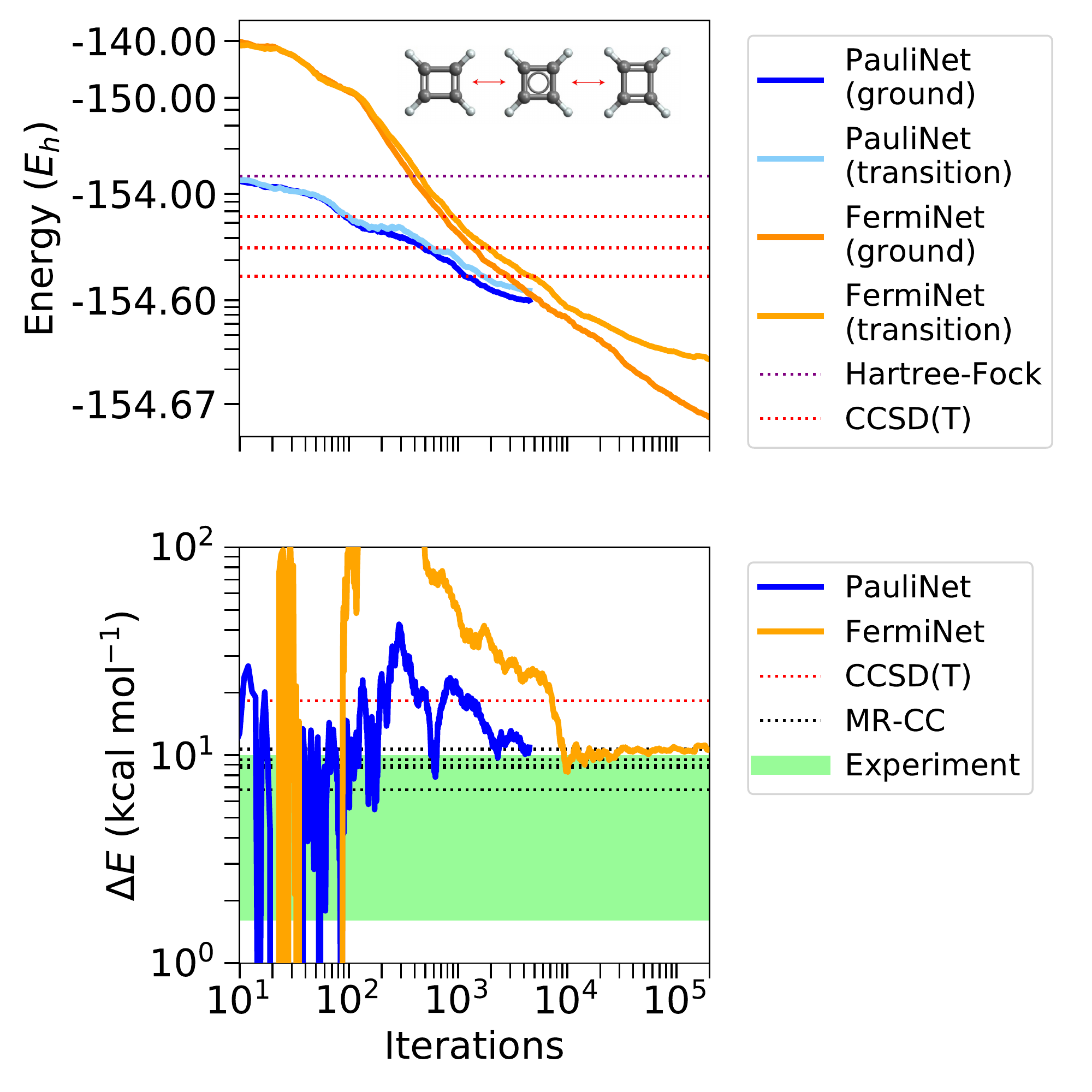}
    \caption{\textbf{Automerization of cyclobutadiene with neural-network ansatzes.}
    Both PauliNet and FermiNet predict relative energies within the range of experimental values and agree with multireference coupled cluster.
    The PauliNet converges more quickly, while the FermiNet reaches lower total energy.
    Figure modified from \citet{Spencer20}.
    }\label{fig:cyclobutadiene}
\end{figure}

\paragraph{Iterative backflow}

\citet{TaddeiPRB15} introduced a form of backflow that applied \cref{eq:traditional-backflow} repeatedly in an interative fashion.
Such an ansatz is formally equivalent to expressing the backflow as a deep neural network \citep{Ruggeri2018-ql}, albeit with artificial restriction on the dimensionality of the hidden layers.
The iterative backflow was used for studying the $^3$He and $^4$He liquids

\paragraph{DeepWF}

The DeepWF \citep{HanJCP19} approach uses an ansatz similar to a Slater--Jastrow wavefunction but with a simpler antisymmetric term:
\begin{equation}
	\psi(\bm{\mathfrak{r}}) = S(\{\bm{\mathfrak{r}}\}, R) A^{\uparrow}(\bm{\mathfrak{r}}^{\uparrow}) A^{\downarrow}(\bm{\mathfrak{r}}^{\downarrow}) .
\end{equation}
The learned symmetric function $S$ is similar to a Jastrow factor and ensures that the wavefunction captures the electron-nuclear and electron-electron cusp conditions.
The antisymmetric factors $A^{\sigma}$ are constructed from the Vandermonde-like determinant of an explicitly antisymmetric two-body function, $A^{\sigma} = \prod_{1\leq i \leq j \leq N} \big(a(\mathbf{r}_i, \mathbf{r}_j, r_{ij}) - a(\mathbf{r}_j, \mathbf{r}_i, r_{ij})\big)$.
The two-body antisymmetric function is entirely learned.
Such a functional form can be evaluated in $\mathcal{O}(N^2)$ operations, compared to $\mathcal{O}(N^3)$ for a determinant.
However, the use of a simplified antisymmetric function is also likely to limit the accuracy achieved: DeepWF obtains only 43.6\% of the correlation energy for the beryllium atom and does not even reach HF accuracy for the boron atom.
The PauliNet and FermiNet approaches described below do much better.
Vanilla PauliNet obtained 99.94\% and 97.3\% of the correlation energies for the beryllium and boron atoms, and FermiNet 99.97\% and 99.83\%, respectively.
Furthermore, FermiNet and PauliNet both substantially surpass conventional Slater-Jastrow-backflow (SJB) wavefunctions on first-row atoms, for which nearly exact benchmark values exist.

\paragraph{PauliNet}

 PauliNet \citep{HermannNC20} builds upon HF or CASSCF orbitals as a physically meaningful baseline and takes a neural network approach to the SJB wavefunction in order to correct this baseline towards a high-accuracy solution (\cref{fig:real_space_networks}a).
 Cusp conditions are explicitly met via the inclusion of cusp correction terms in the wavefunction \citep{Ma2005-cusps}.
 A graph-convolutional block based on SchNet \citep{schutt2018schnet} is used to create a permutation-equivariant latent space representation depending on the many-electron configuration.
 This embedding is then passed into separate deep neural networks that learn the Jastrow factor and a (cuspless) backflow transformation.
 \citet{HermannNC20} introduced PauliNet with a purely multiplicative backflow as shown in \cref{fig:real_space_networks}a; \citet{SchatzleJCP21} generalized this to a multiplicative and additive backflow as shown in \cref{eq:backflow_mult_add}.
 PauliNet is optimized with a fixed number of Slater determinants.
Most of the results reported in \citet{HermannNC20,SchatzleJCP21} were obtained with around 10 determinants.

\paragraph{FermiNet}

FermiNet \citep{PfauPRR20} takes a more minimalist (or machine-learning maximalist) approach and attempts to train a neural network to represent the entire wavefunction (\cref{fig:real_space_networks}b).
FermiNet uses two parallel networks, describing one- and two-electron features respectively.
The inputs to each layer in the one-electron stream are permutation-equivariant functions of the activations from the previous layers of the one- and two-electron streams.
The final layer projects the latent space into the required number of orbitals, from which determinants can be formed and evaluated.
As with PauliNet, the final wavefunction is a sum over a number of  determinants.
For most of the results reported in \citet{PfauPRR20}, 16 determinants were used.
FermiNet builds up a rich description of electron-electron interactions from the permutation-equivariant mixing of information describing one- and two-electron features.
In particular, the electron-nuclear and electron-electron cusps in the wavefunction are represented accurately, despite not being encoded explicitly.
Whereas PauliNet is usually trained with the ADAM optimizer, FermiNet training was found to be substantially improved when employing the KFAC optimizer.

While both PauliNet and FermiNet exceed the accuracy of conventional SJB wavefunctions on small systems, there are important tradeoffs between the two models.
Results from both on the automerization of cyclobutadiene can be seen in \cref{fig:cyclobutadiene}.
The FermiNet is typically trained with a larger number of parameters than the PauliNet, requiring more iterations and more computation per iteration to converge, but it typically converges to a lower absolute energy.
Recently, \citet{gerard2022gold} proposed a hybrid ansatz which uses neural network layers similar to the SchNet and PauliNet in a FermiNet-like architecture.
This hybrid ansatz was found to reach even lower absolute energies than the FermiNet on systems like benzene and the potassium atom.

\paragraph{Potential energy surfaces}

Typically one optimises a wavefunction at a specific geometry but this quickly becomes prohibitively expensive for exploring the high-dimensional potential energy surface of even relatively small molecules.
\citet{ScherbelaNCS22} developed a training methodology that allows weight sharing between (simplified) PauliNet architectures targeting different geometries.
By switching the geometry being trained at each epoch, they showed that the computational cost for training across a set of geometries can be improved by an order of magnitude without affecting the accuracy of the final energies, with 95\% of network parameters shared across all geometries.
This implies that the network is learning features of electron correlation in general rather than fitting to a specific geometry.
They also demonstrated that a wavefunction for a larger molecule could be initialised from a wavefunction for a smaller molecule and could then be fine-tuned in a relatively short optimization stage.
Pretraining neural network wavefunctions from smaller systems has also been shown to dramatically accelerate convergence for Kagome lattice models \citep{Yang2020-bk}.

In a similar vein, \citet{Gao2021-cg, gao2022sampling} demonstrated that a meta-learning approach, where a graph neural network is used to parameterize a wavefunction model, can accurately represent the wavefunctions for multiple geometries, enabling a fully quantum-mechanical potential energy surface to be represented in a single model.
Their approach used a FermiNet-like wavefunction model, but the meta-learning concept is directly applicable to other wavefunction representations, assuming the wavefunction form is sufficiently flexible.

\paragraph{Periodic systems}

There has also been progress on using first-quantized neural network architectures in periodic systems, such as interacting quantum gases in low dimension \citep{pescia_neural-network_2022}, the electron gas \citep{wilson2022-ueg,cassella2022-ueg,Li2022-abinitio}, and for small cells of solids such as lithium hydride and graphene \citep{Li2022-abinitio}.
Again, sufficiently expressive networks at the VMC level have been found capable of rivalling or surpassing the accuracy of fixed-node diffusion Monte Carlo calculations using conventional Slater-Jastrow-backflow trial wavefunctions.

\subsection{Extensions}

\paragraph{Pseudopotentials}

The electronic structure of heavy atoms, especially transition metals, is complicated and challenging for all QC methods.
The difficulty is compounded by the high computational cost of variational Monte Carlo methods, which scale roughly as $\mathcal{O}(Z^{5})$ \citep{Hammond1987}, where $Z$ is the nuclear charge.
Whilst the core electrons contribute heavily to the total energy, energy differences are largely determined by the behaviour of the valence electrons.
The core electrons can therefore be removed and the effective nuclear charge reduced by the use of pseudopotentials.
The use of pseudopotentials is common in many methods, including density functional theory and conventional variational Monte Carlo.
\citet{Li2022-pseudo} demonstrate that effective core potentials can be readily combined with FermiNet and achieve accuracy comparable to CCSDT{(Q)} extrapolated to the complete basis set limit for first-row transition metal atoms.
The computational time per iteration was reduced by 43\% (17\%) for the scandium (zinc) atom using an argon core.
Again, this approach is not restricted to FermiNet.
Pseudopotentials can be used with any first-quantized neural network wavefunction.

\paragraph{Diffusion Monte Carlo (DMC)}

Projector methods such as DMC \citep{needs2020variational} and auxiliary-field Monte Carlo \citep{ShiJCP21} go beyond VMC by using stochastic algorithms to sample the ground state without requiring its wavefunction to be represented as a known function or network.
DMC is in principle exact but, for many-fermion systems, relies in practice on the fixed-node approximation, in which collapse to the bosonic ground state is avoided by imposing the sign structure of the trial wavefunction on the DMC wavefunction.
A DMC simulation therefore samples (stochastically) the lowest energy state with the same sign structure as the trial wavefunction.
The improvements that result from applying DMC to conventional Slater-Jastrow-backflow trial functions optimized using VMC methods are substantial, explaining why DMC is so often used to provide improved estimates of the ground-state wavefunction and energy.
\citet{Wilson21} combined DMC with a FermiNet trial wavefunction.
For first-row atoms, DMC captured much of the remaining correlation energy (94\% of the difference between the VMC energy and the exact energy in the case of the nitrogen atom).
However, \citet{Wilson21} used a simplified FermiNet that gave VMC energies higher than those reported by \citet{PfauPRR20}, which were already within 1mH of exact results for all first-row atoms.
Given evidence that the mean-field equivalent of PauliNet can essentially match HF in the complete basis set limit \citep{SchatzleJCP21}, it is possible that the remaining error in PauliNet and FermiNet wavefunctions is dominated by errors in the nodal surface, which are rarely sampled regions during optimisation.
If this is the case, diffusion Monte Carlo with the fixed node approximation may not produce substantially lower energies.
On the other hand, since neural network wavefunctions routinely capture over 90\% of the correlation energy at the VMC level, the need to perform expensive diffusion Monte Carlo calculations is greatly reduced.
More recently, \citet{Ren22} showed that DMC can capture roughly half of the remaining correlation energy for the atoms Li-Ar, when using a very small FermiNet-based architecture.
Whilst it is possible to achieve energies within chemical accuracy using FermiNet at the VMC level, these calculations model the case for larger systems where converging the energy with respect to network size might not be feasible.
\citet{Ren22} went on to demonstrate that DMC using FermiNet trial wavefunctions noticeably reduces the energy for larger systems.
In the case of the benzene dimer, the reduction was 50mH.

\paragraph{Excited States}

Our discussion so far, and most VMC calculations, have focused on ground state properties.
However, excited states are of critical importance to understand the behaviour of materials.
Fortunately, recent algorithmic developments by multiple groups have demonstrated that the calculation of excited states using VMC methods is feasible and can achieve an acceptable trade-off in accuracy and cost.
Here we highlight three such approaches utilizing conventional VMC wavefunctions.
One approach is the state-averaged VMC method \citep{Schautz2004,Dash2019}, in which the average energy over multiple states is minimised and individual states are projected out via diagonalization within the basis of excited states.
Similar techniques are used with other quantum chemistry methods.
\citet{Zhao2016} instead minimized a different objective function, such that the state with energy closest to a desired energy target is obtained.
\citet{Pathak2021} suggested a simple alternative, where a state is forced to be (approximately) orthogonal to all lower energy states via a penalty term.
These techniques can be readily applied to VMC using neural-network wavefunctions and, in particular, penalty function approaches have recently been explored.
As with ground-state calculations, the flexibility of the wavefunction ansatz to represent the desired state is critical.
\citet{Entwistle22} demonstrated that the PauliNet architecture combined with a penalty function  can represent the lowest few excited states of molecules up to the size of benzene (\cref{fig:real_space_networks}c).
Relatedly, \citet{ChooPRL2020}  demonstrated that NQS on lattice models can obtain the lowest-energy state of any given Abelian symmetry by performing what is essentially a ground-state simulation in that symmetry sector, and multiple states of the same symmetry using a penalty function.
However, the most accurate and efficient way to obtain excited states within VMC, irrespective of wavefunction ansatz, remains an open question \citep{Cuzzocrea2020}.

\section{Electrons in second quantization}
\label{sec:discrete-space}

\begin{figure*}[t]
\centering
\includegraphics[width=\textwidth]{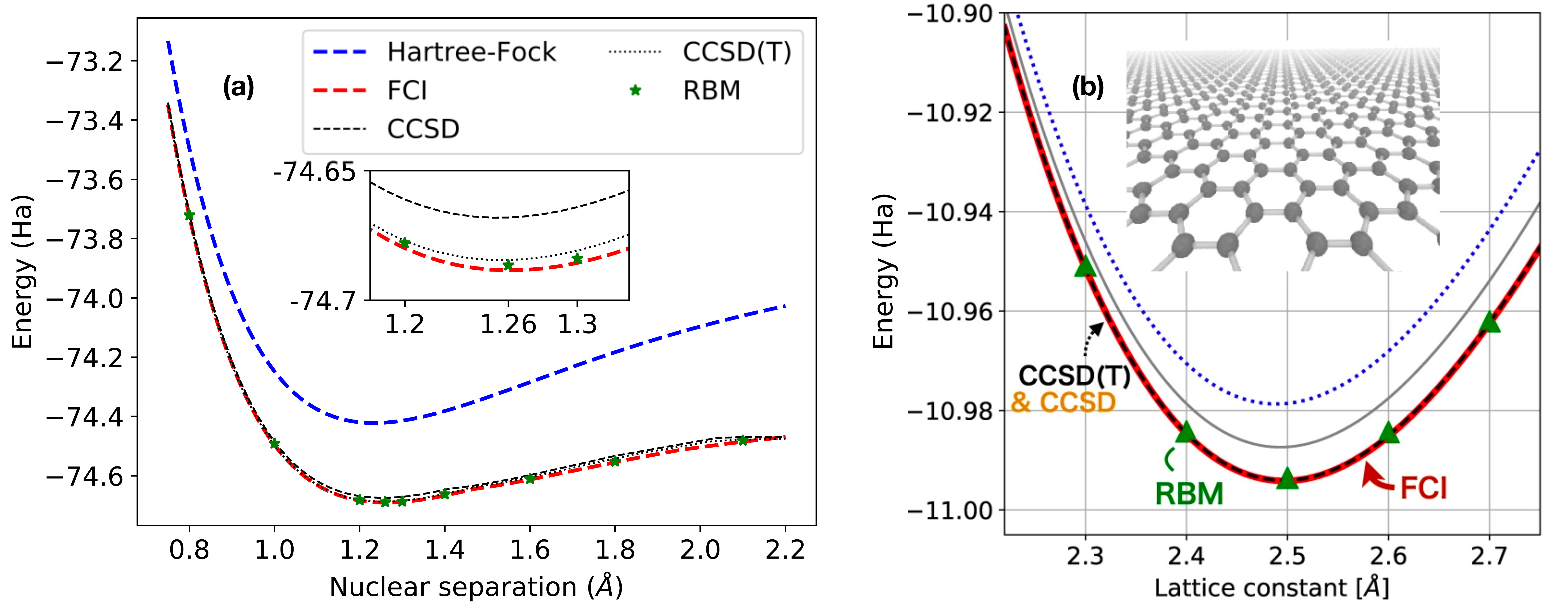}
\caption{\textbf{Electronic energies for molecules and solids in second quantization}
(\textbf a) Dissociation curve for $C_2$ molecule in the STO-3G basis.
The green stars show results for a restricted Boltzmann machine which represents the electrons in discrete space.
Figure taken from \citep{ChooNC20}.
(\textbf b) Graphene on a honeycomb lattice solved using the cc-pVDZ basis set.
Figure taken from \citep{yoshioka2021solving}.
}\label{fig:discrete-space}
\end{figure*}

Instead of working directly with the infinite-dimensional Hilbert space corresponding to the real-space Hamiltonian of \cref{eq:hamiltonian}, it is common practice in QC to use a finite basis set.
By choosing a set of electronic basis functions $\{\varphi_{1}(\mathbf{r}), \varphi_{2}(\mathbf{r}), \dots\}$, we can define a set of second-quantised operators $\hat c^{\dagger}_{i}$ ($\hat c_i$) which create (annihilate) an electron in the $i$-th basis function, and which satisfy the canonical anticommutation relations $\{\hat c^\dag_i, \hat c_j\} = \delta_{ij}$.
These operators then act on the second-quantized wavefunction $\psi_{n_1n_2\cdots}$, which encodes amplitudes for different occupations of the orbitals (Box~\ref{box:first-second-quant}).
Projecting the real-space Hamiltonian onto this set of orbitals then yields the corresponding discretized Hamiltonian,
\begin{equation}
\label{eq:2quantised Hamiltonian}
\hat H=\sum_{ij} t_{ij} \, \hat c^\dag_{i} \hat c^{\phantom{\dag}}_{j} +  \sum_{ijkl} u_{ijkl}\, \hat c^\dag_{i} \hat c^\dag_{j} \hat c^{\phantom{\dag}}_{l} \hat c^{\phantom{\dag}}_{k} ,
\end{equation}
where
\begin{align}
t_{ij} &= \int \varphi_i^{\ast}(\mathbf{r}) \left ( -\frac{1}{2}\nabla^2 - \sum_I \frac{Z_I}{|\mathbf{r} - \mathbf{R}_I|} \right ) \varphi_j(\mathbf{r}) \, d\mathbf{r} , \\
u_{ijkl} &= \iint \varphi_i^{\ast}(\mathbf{r}) \varphi_j^{\ast}(\mathbf{r}') \frac{1}{|\mathbf{r} - \mathbf{r}'|} \varphi_k(\mathbf{r}) \varphi_l(\mathbf{r}') \, d\mathbf{r} d\mathbf{r}' ,
\end{align}
are matrix elements of the one- and two-electron terms in the real-space Hamiltonian of Eq.~\eqref{eq:hamiltonian}.
For simple basis functions such as Gaussians or plane waves, the matrix elements can be evaluated analytically.
This Hamiltonian serves as the starting point for the methods described in this section.

\subsection{Fermionic neural quantum states}\label{sec:second_quantization}

Instead of working directly with the occupation-number representation of the wavefunction (Box~\ref{box:first-second-quant}), it is also possible to map occupation numbers $n_k\in\{0,1\}$ onto degrees of freedom $\sigma^z_k\in\{\downarrow,\uparrow\}$ of spin-1/2 particles, such that empty orbitals map to down spins and occupied orbitals to up spins.
This mapping makes it possible to leverage NQS and other methods for solving quantum spin systems.
The same duality allows the creation and annihilation operators appearing in the electronic Hamiltonian (\pcref{eq:2quantised Hamiltonian}) to be written in terms of spin operators.
This can be achieved, for example, with the Jordan--Wigner mapping \citep{Wigner1928}, that transforms annihilation and creation operators into, respectively, lowering and raising spin operators $\hat \sigma^\pm_j = (\hat \sigma^x_j  \pm i \hat \sigma^y_j)/2$.
This mapping is not unique, however, and there exist more recent alternatives, such as parity or Bravyi--Kitaev encodings \citep{BK2002}, both of which have been developed in the context of quantum simulations.

Regardless of the choice of spin encoding, the final outcome is a spin Hamiltonian with the general form
\begin{equation}
\label{Hpauli}
\hat H_{q} = \sum_{p=1}^{r} h_p \hat{\Xi}_p ,
\end{equation}
defined as a linear combination with real coefficients $h_p$ of $\hat{\Xi}_p$, which are
$N$-fold tensor products of single-qubit Pauli operators and the identity:  $\hat I,\hat \sigma^x,\hat \sigma^y,\hat \sigma^z$.

The ground state of the spin Hamiltonian in \cref{Hpauli} can be approximated using a spin-based NQS representation based on complex-valued RBMs \citep{CarleoS17}.
For a system of $N$ spins, the many-body amplitude corresponding to a state in the $\sigma^z$ basis, i.e., $\boldsymbol{\sigma} = (\sigma^z_1 \dots \sigma^z_N$), takes the compact form
\begin{equation}
\label{RBM}
\psi(\boldsymbol\sigma; \boldsymbol{\theta}) = e^{\sum_i a_i \sigma^z_i} \prod_{j=1}^M 2 \cosh
\biggl(b_j + \sum_i^N W_{ij} \sigma^z_i\biggr) ,
\end{equation}
with parameters $\boldsymbol\theta = {(a_i,b_j,W_{ij})}$.
This ansatz can be optimised with VMC techniques (Box~\ref{sec:optimization}), typically relying on the stochastic reconfiguration \citep{SorellaPRL98} approach.
A number of works have adopted this approach and achieved competitive variational results for small basis sets \citep{ChooNC20, YangJCTC20}, even in conjunction with quantum computers \citep{torlai2020precise,iouchtchenko2022neural}.
In \cref{fig:discrete-space} (a), we show the dissociation curve of $C_{2}$, in the STO-3G basis, using the RBM as described above \citep{ChooNC20}.

\paragraph{Solids}

The second-quantization framework also allows one to treat solids, using as a basis the Bloch orbitals obtained by solving the crystalline HF equations \citep{del_re_self-consistent-field_1967}.
Creation and annihiliation operators, $\hat c^{\dagger}_{i \mathbf{k}}$ and $\hat c_{i \mathbf{k}}^{\phantom{\dagger}}$, for electrons in band $i$ with crystal momentum $k$ are introduced, and the resulting Hamiltonian is similar to \cref{eq:2quantised Hamiltonian}, with the noticeable difference that the one- and two-body matrix elements now depend on the crystal momenta: $t_{ij}^{\phantom{k}} \rightarrow t^{\mathbf{k}}_{ij}$ and $u_{ijkl}^{\phantom{\mathbf{k}_1}} \rightarrow u^{\mathbf{k}_1\mathbf{k}_2\mathbf{k}_3\mathbf{k}_4}_{ijkl} $, with the four momenta appearing in the two-body integrals satisfying the conservation of the total crystal momentum.
Using Gaussian-based atomic functions as the single-particle basis and RBM wavefunctions to represent the many-body state, \citep{yoshioka2021solving} applied this approach to study the electronic structure of solids.
In \cref{fig:discrete-space} (b), we show the computed ground-state energies for graphene crystals as a function of the lattice constant.

\paragraph{Exact Sampling}

Fermionic NQS are typically sampled using the MCMC approach commonly adopted in VMC (Box~\ref{sec:QMC}).
However, the mixing rate of the MCMC algorithm is known to be slow in some cases, such as close to phase transitions, and MCMC simulations can suffer from critical slowing down.
A way to circumvent this limitation is to introduce model wavefunctions explicitly designed to allow exact sampling of their square modulus, thus avoiding the need to use MCMC\@.
One such family are autoregressive neural network wavefunctions \citep{sharir2020deep}, a complex-valued generalization of the autoregressive models commonly adopted in deep learning.
Such networks represent normalized wavefunctions and allow one to directly obtain perfectly uncorrelated samples; this is useful as the wavefunction distribution for many QC problems can be highly multi-modal.
The exact sampling approach was applied to QC hamiltonians in a recent work by  \citet{BarrettNMI22}.
Optimizations in the way Hamiltonian matrix elements and the corresponding Monte Carlo estimators are computed have made it possible to treat much larger systems than were accessible in the early applications of \citet{ChooNC20}.
Specifically, \citet{zhao_scalable_2022}) obtain  competitive variational energies, improving on the CCSD energies of molecules in minimal basis sets.
Results for up to around 50 electrons in 80 orbitals (\ce{Na2CO3} at equilibrium) have been obtained at relatively modest computational cost.

\subsection{ML-assisted selected CI}

For many QC problems, although the dimension of the Hilbert space grows exponentially with system size, the number of relevant configurations in the ground state typically remains sparse.
This suggests that by efficiently selecting the relevant configurations and then diagonalising the Hamiltonian on the reduced subspace, one can achieve highly accurate results.
This set of approaches is also known as selected CI \citep{HuronJCP73, giner2013using, holmes2016heat, sharma2017semistochastic}.
Different flavours of selected CI vary in the way relevant configurations are selected.

One well-known approach is called Monte Carlo CI (MCCI) \citep{greer1998monte} and can be briefly summarised as follows:
\begin{enumerate}
    \item Start from a finite set of configurations $S_i = \lbrace \ket{x} \rbrace$
    \item By considering single or double excitations starting from configurations in $S_i$, construct an expanded set $S_{i}'$.
    \item Construct the Hamiltonian $\hat H_i$ for the expanded set $S_{i}'$ and diagonalise to obtain the wavefunction coefficients for the configurations in the set.
    \item Discard the configurations whose coefficient is less than a given threshold $c_{\text{min}}$.
The remaining configurations then form a new set of configurations $S_{i+1}$.
    \item Repeat until convergence.
\end{enumerate}
ML techniques can be used to improve selection of the configuration set.
One such approach is to perform supervised learning \citep{Coe2018Machine, GlielmoPRX20}, where a neural network is trained to predict the wavefunction coefficients using the data from the MCCI method, i.e., the wavefunction coefficients of the configurations in the set $S_{i}'$.
After training, the network can be queried or sampled to select the configurations with the largest coefficients.
In other words, the network is used to bootstrap and predict the coefficients of configurations not yet seen in the data set.
It was shown in \citet{Coe2018Machine} that such an approach converges faster than the vanilla MCCI method.

The task of selecting configurations for selected CI can also be cast as a reinforcement-learning task where the state is the current set of configurations and an agent is trained to perform actions on the set to iteratively modify the configurations with the aim of minimising the variational energy.
This approach was applied in \citet{Goings2021Reinforcement} to achieve near-FCI accuracy for small molecules in a small basis set.

\section{Challenges and outlook}
\label{sec:conclusion}

Ab-initio QC with neural-network wavefunctions has only just emerged as a viable path to highly accurate electronic-structure methods, yet it already competes with established approaches that have been developed for decades.
We imagine that it may become the methodology with the best trade-off between efficiency and accuracy for systems with up to one to two hundred electrons and a nontrivial electronic structure.
Before that can happen, however, several challenges must be addressed.
All the methods are currently in a development stage and only limited benchmarking is available.
As such, it is not yet clear whether the excellent accuracy seen so far will be maintained across a broader range of chemical systems, or how rapidly the accuracy will degrade with system size.
Related to this is our incomplete understanding of what limits the accuracy of neural-network ansatzes, and how their success or failure is related to physical phenomena such as strong correlation.
Since the underlying electronic problem is exponentially hard but the algorithms are polynomial, they must be limited in accuracy in some ways.
It is not currently clear, however, whether the limitations seen to date are caused by the restricted expressiveness of the neural networks or by difficulties in optimization or both.
For instance, while it has been proven that a single generalized Slater determinant is in principle sufficient to represent any antisymmetric function \citep{Hutter20}, it might not be possible to parametrize it with a polynomially scaling neural network or train it within a polynomially scaling time.

Apart from these fundamental issues, there are many practical challenges.
While the scaling of variational QMC with system size is favourable, the prefactor due to the neural networks is large.
Until very recently, this limited applications to systems no larger than the benzene molecule (42 electrons), which is three to four times below our envisaged applicability range, although results for a 108-electron simulation cell of solid LiH have now been reported \citep{Li2022-abinitio}.
The prefactor can be reduced by integrating traditional QC techniques such as pseudopotentials \citep{Li2022-pseudo}, developing more efficient neural-network architectures, or using ML techniques such as pre-training and transfer learning.
Specific to the discrete-basis second-quantized approaches is the issue of basis-set convergence, where sufficiently large basis sets may increase the prefactor by up to three orders of magnitude compared to minimal basis sets.
Another challenge is related to the stochastic optimization, which produces noise in the converged energies that is especially amplified when calculating small energy differences.

We are, however, optimistic that many of these challenges can be addressed and can be addressed quickly, thanks to the relative simplicity of the framework based on variational QMC and of neural networks compared to traditional QC approaches.
Indeed, this simplicity has already enabled rapid development of multiple extensions to the first single-point ground-state calculations on molecules, including transferable wavefunctions, excited states, and formulations for periodic systems, all originating from multiple independent research groups.

First-quantized approaches such as FermiNet, PauliNet, and their successor architectures already match essentially exact benchmark results to within chemical accuracy for small systems.
Yet these networks are just a small subset of possible architectures for representing antisymmetric wavefunctions, and it is unlikely that the optimal ones were found on the first attempt, so we expect that significant innovation lies ahead.
We believe that ab-initio methods based on neural-network wavefunctions will become an integral part of the QC toolbox that enables straightforward electronic-structure calculations of complex molecular systems.

\begingroup
\setlength\bibitemsep{0pt}
\printbibliography%
\endgroup

\subsection*{Acknowledgements}

\begingroup
\footnotesize

We acknowledge funding from the German Ministry for Education and Research (Berlin Institute for the Foundations of Learning and Data, BIFOLD), the Berlin Mathematics Research Center MATH+ (AA1-6, AA2-8), and European Commission (ERC CoG 772230 ScaleCell).
\endgroup

\end{document}